\documentclass[a4paper,
               keeplastbox,   
               ]{jacow}
\usepackage{pdfpages,multirow,ragged2e} %
\makeatletter%
	\ifboolexpr{bool{xetex}}
	 {\renewcommand{\Gin@extensions}{.pdf,%
	                    .png,.jpg,.bmp,.pict,.tif,.psd,.mac,.sga,.tga,.gif,%
	                    .eps,.ps,%
	                    }}{}
\makeatother

\ifboolexpr{bool{xetex} or bool{luatex}} 
 {}                                      
 {\usepackage[utf8]{inputenc}}           

\usepackage[USenglish]{babel}
\usepackage{float}

\ifboolexpr{bool{jacowbiblatex}}%
 {%
  \addbibresource{jacow-test.bib}
  \addbibresource{biblatex-examples.bib}
 }{}
\begin{document}

\title{POSITRON DRIVEN MUON SOURCE FOR A MUON COLLIDER \thanks{Work supported by INFN, Frascati National Laboratories }}

\author{
D. Alesini, M. Antonelli, M.E. Biagini, M. Boscolo, O.~R.~Blanco-Garc\'ia, A. Ciarma,\\
        R. Cimino, M. Iafrati, A. Giribono, S. Guiducci, L. Pellegrino, M. Rotondo,\\ C. Vaccarezza, A. Variola\thanks{e-mail address: alessandro.variola@lnf.infn.it}, INFN-LNF, 00044 Frascati, Italy \\
		A. Allegrucci, F. Anulli, M. Bauce, F. Collamati, G.Cavoto,\\ G. Cesarini, F. Iacoangeli, R. Li Voti, INFN-Roma, 00185 Roma, Italy  \\
		A. Bacci, I. Drebot, INFN-MI, 20133 Milano, Italy \\
		P. Raimondi, S. Liuzzo, ESRF, 38043 Grenoble, France\\
        I. Chaikovska, R. Chehab, IN2P3-LAL, 91440 Orsay, France\\
        N. Amapane, N. Bartosik, C. Bino, A. Cappati, G. Cotto,\\ N. Pastrone, M. Pelliccioni, O. Sans Planell INFN-TO, 10125 Torino, Italy\\
         M. Casarza, E. Vallazza, INFN-TS, 34127, Trieste, Italy\\
        G. Ballerini, C. Brizzolari, V. Mascagna, M. Prest,\\ M. Soldani, Insubria University, 22100 Como, Italy\\
        A. Bertolin, C. Curatolo, F. Gonella, A. Lorenzon, D. Lucchesi,\\ M. Morandin, J. Pazzini, R. Rossin, L. Sestini, S. Ventura,\\ M. Zanetti, Padova University, 35121 Padova, Italy and INFN-PD, Padova, Italy\\
        F. Carra, P. Sievers, CERN, CH-1211 Geneva 23, Switzerland\\
        L. Keller, SLAC National Accelerator Laboratory, 94025 Menlo Park, CA, US\\
        L. Peroni, M. Scapin, Politecnico di Torino, 10129 Torino, Italy\
        }

\maketitle

\begin{abstract}
   The design of a future multi-TeV muon collider needs new ideas to overcome the technological challenges related to muon production, cooling, accumulation and acceleration. In this paper a layout of a positron driven muon source known as the \textbf{L}ow \textbf{EM}ittance \textbf{M}uon \textbf{A}ccelerator (LEMMA) concept (see refs. \cite{Snowmass, PRAB, NIM, rast10}) is presented. The positron beam, stored in a ring with high energy acceptance and low emittance, is extracted and driven to a multi-target system, to produce muon pairs at threshold. This solution alleviates the issues related to the power deposited and the integrated Peak Energy Density Deposition~(PEDD) on the targets. Muons produced in the multi-target system will then be accumulated before acceleration and injection in the collider. A multi-target line lattice has been designed to cope with the focusing of both the positron and muon beams. Studies on the number, material and thickness of the targets have been carried out. A general layout of the overall scheme and a description is presented, as well as plans for future R\&D.
\end{abstract}

\section{INTRODUCTION}
In this paper we will describe a scheme  to produce
low emittance muon beams from 
electron-positron collisions at centre-of-mass energy just above the $\mu^{+}\mu^{-}$  production
threshold with maximal beam energy asymmetry, that corresponds to about 45 GeV
positron beam interacting on an electron target. Previous studies on this subject are reported in ref.~\cite{Snowmass, PRAB, NIM}.\par

The most important key properties of the muons produced by the $e^+$ on target are:
the low and tunable muon momentum in the centre of mass frame, and the large boost, being about $\gamma\sim$200. These characteristic results in the following advantages: the final state muons are highly collimated and have small emittance, not requiring any cooling stage, and they have an average laboratory lifetime of almost 500 $\mu$s.

As far as the muon bunch intensity is concerned, the number of $\mu^{+}\mu^{-}$ pairs produced per positron bunch on target is:
\begin{equation}
\label{eq:nmu}
  n(\mu^{+} \mu^{-})=n^{+} \rho^{-} l \sigma(\mu^{+} \mu^{-})      
\end{equation}
where $n^{+}$ is the number of $e^+$ in
the bunch, $\rho^-$ is the electron density in the medium, $l$ is
the thickness of the target, and $\sigma(\mu^{+} \mu^{-})$ is the muon pairs production cross section.
 
\begin{figure}[htbp!]
  \centering
  \includegraphics[width=\columnwidth]{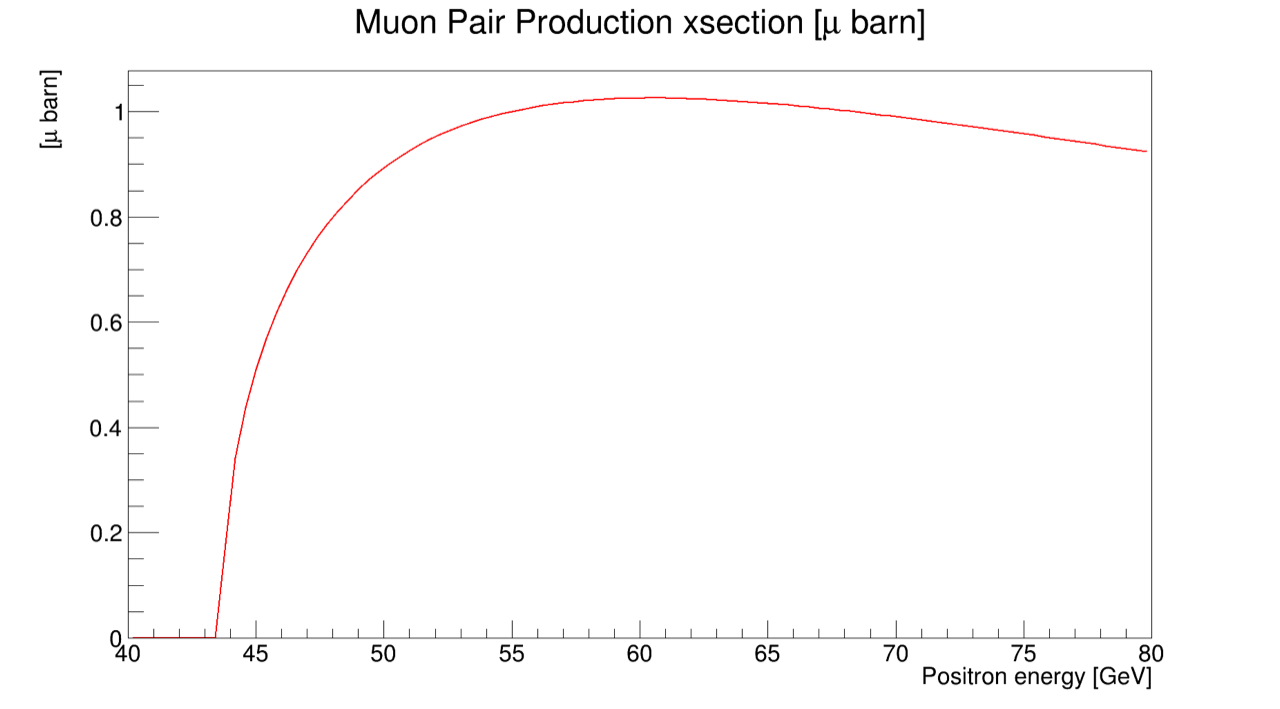}
  \caption{$e^{+}e^{-}\rightarrow\mu^{+}\mu^{-}$ cross section as a function of the positron beam energy ($\sqrt{s} \sim$0.23 GeV = 53 GeV positron beam).
    \label{fig:xsec}}
\end{figure}

The cross section for continuum muon pair production $e^{+}e^{-}\rightarrow\mu^{+}\mu^{-}$ just above threshold (see Fig.~\ref{fig:xsec}) approaches its maximum value of about 1$\mu$b at $\sqrt{s} \sim$0.23 GeV. This requires a target with very high electron density to obtain a reasonable muon production efficiency. Such high-density values can be obtained
either in a liquid or solid target or, possibly, in a more exotic
solution like in crystals. 

Collinear radiative Bhabha scattering, with a cross section of about of 150 mb, 
actually sets the value of the positron beam interaction length for a given pure electron target density value. Using as reference value for the positron beam degradation when its current is decreased by 1/$e$, i.e. one beam lifetime, one can
determine the maximum achievable value for the target density and length:$
(\rho^- l)_{max} =  1/\sigma(rad.bhabha)\approx 10^{25}  \hbox{cm}^{-2}$.
The ratio of the muon pair production cross section to the radiative bhabha cross section determines the maximum value of the {\it muons conversion efficiency} $eff(\mu^+\mu^-)$, defined as the ratio of the number of produced $\mu^+\mu^-$ pair to the number of the incoming $e^+$.
Easily one can see that  the upper limit of $eff(\mu^+\mu^-)$  is of the order of $10^{-5}$, so that: $n(\mu^+ \mu^-)_{max}\approx n^+ 10^{-5}$.\par

Electromagnetic interactions with nuclei (bremmstrahlung, and multiple Coulomb scattering) are dominant in conventional targets, then in absence of intrinsic focusing effects, a target thickness increase corresponds to an increase in the muons beam emittance $\epsilon_{\mu}$.
In this framework positrons and muons production and interaction on different targets, liquid Hydrogen, Liquid Lithium, Beryllium, and Carbon, Diamond and Copper and heavier materials, have been studied in \cite{PRAB, NIM} with {\texttt{GEANT4}}\cite{geant} and FLUKA.

For all cases, the target thickness and the positron beam energy have been optimized to maximize key parameters. As expected, it has been found that light materials, as liquid  Hydrogen, Beryllium, Carbon, and Diamond, have a better performance
with respect to heavier materials (i.e. Copper), having a larger muon production efficiency $eff(\mu^+\mu^-)$. In addition in these cases the muon beam is produced with a smaller emittance.  \par

As a further development in Ref. \cite{NIM} two conceptual schemes were proposed: the single pass and the multi--pass of positron bunches on target. In both cases the low value of the muon conversion efficiency requires two Muon Accumulator (MA) rings to reach $O(10^8)$ $\mu^+$ and $\mu^-$ per bunch. The muon bunches can be stacked in the same phase space intercepting the positron beam in the interaction point where the muons are generated in targets.
The muon laboratory lifetime $\tau_{\mu}^{lab}$ is about 460 $\mu$s so that the stacking of the muon bunches need to be fast. 
In particular, a multi--pass scheme has been studied in Ref. \cite{PRAB}, implemented in a large momentum acceptance storage ring, so allowing for increasing the $\mu$ conversion efficiency. In this scheme, limitations on ring energy acceptance, thermo--mechanical stress on target, and muon recombination all required very challenging innovations.\par

The next mandatory step was to implement conceptual schemes into a consistent accelerator complex design, in order to have a baseline layout demonstrating the feasibility of the expected parameters of all the sub-systems which compose the LEMMA muon source.
In this framework different considerations can be taken into account. The solutions will depend on the single sub-system design, the possible R\&D effort to be performed, and the final required parameters.
To demonstrate the consistency of the design, taking into account the main limitations of the LEMMA scheme, i.e. the positron production rate, the low efficiency of muon production, the high synchrotron power in the 45 GeV Positron Ring, and the short muon lifetime at production, three different schemes were studied and the full working cycle for each elaborated.

\section{MUON SOURCE SCHEMES}
To elaborate a scheme for the LEMMA muon source it is necessary to start from the main technical constraints that are imposed by the muon physics and the technological limits. The full muons production cycle should be less than the about 410 $\mu$sec given by a fraction of the single particle lifetime (467 $\mu$sec) at 22.5 GeV, thus reducing the intrinsic beam losses with respect to the accumulated intensity. After the production cycle the bunches must be immediately re-accelerated to increase the lifetime and freeze the losses. Moreover the full cycle must accommodate enough time for the positron source production and cooling, either in the main positron ring or in a dedicated damping ring. This damping time must be compatible with a reasonable amount of synchrotron power emitted so it can range from 10 msec in a low energy damping ring to 80 msec in a high energy positron ring. This also in case it is possible to recuperate a part of the positron bunches spent in the muon production that, after the targets interactions, are strongly affected and their 6D emittance degraded. \par

It is then evident that the impact of the muon production on the $e^+$  bunches should be minimized to allow generating the maximum amount of muons for a single $e^+$  bunch passage and this can be done evaluating different type of targets. Once a $e^+$ bunch has been spent it is mandatory to take into account a new $e^+$ bunch for the muon accumulation cycle, the number of ``fresh'' $e^+$ bunches available in an high energy positron ring being so another physical limit. This limit impose a very large $e^+$ storage ring design to accommodate the maximum of ``fresh'' $e^+$ bunches without drastically increasing the average current and so the beam power; as classical example, in this first phase, the LHC circumference of 27 km has been considered. Furthermore the different systems composing the source complex must show not unrealistic performances taking into account the state of the art of the existing technology or the possibility to have future solid R\&D program to fulfill the required parameters. In this framework it is important to highlight the critical aspect represented by the $e^+$ source that is supposed to provide performances more than three order of magnitude in respect to the existing one (SLAC).
All these considerations have been considered as the basis for the proposal of three different schemes representing the first step towards a pre-conceptual design, taking into account a baseline configuration and some alternatives. These schemes have been studied considering the layout of the different systems composing the muon source but also a complete cycle description based on all the different muons production phases. The repetition rate for Scheme I and II is 10 Hz, while for Scheme III is 20 Hz.

\section{Scheme I }
\subsection{General layout}
This first scheme is based on the 27 km $e^+$ Positron Ring (PR), where 1000 bunches with $5x10^{11}$ $e^+$/bunch are stored at 45 GeV. The $e^+$ bunches are extracted from the ring and sent to a multiple targets straight section, where muons are produced. After production, the $\mu^{+}$ and $\mu^{-}$ bunches are re-circulated in separate rings sharing the production line, where they will circulate synchronously with the incoming ``fresh'' $e^+$ bunch in the targets line, thus increasing their phase space density up to a final availability of two bunches of $\sim$10$^9$ muons.
In this Scheme, an $e^+$ compressor Linac at the exit of the muon production line will be needed in order to compress the longitudinal energy spread before the re-injection of the degraded post-production $e^+$ beam in the PR. The goal is to allow for recuperating, without losses, at least 90\% of the original $e^+$ bunch intensity, so drastically reducing the $e^+$ source requirements of one order of magnitude. 
The $e^+$ source can be split in two parts each contributing in restoring the required $e^+$ bunch population in the PR:
\begin{itemize}
    \item a classical $e^+$ source based on amorphous, crystal or hybrid target is also providing the first slow spill injection in the main ring,
    \item due to the high gamma rays flux produced in the muon production target it is possible also to envisage another embedded source based on an amorphous target at the end of this line.
\end{itemize}

\begin{figure}[htpb]
    \centering
	\includegraphics[width=\columnwidth]{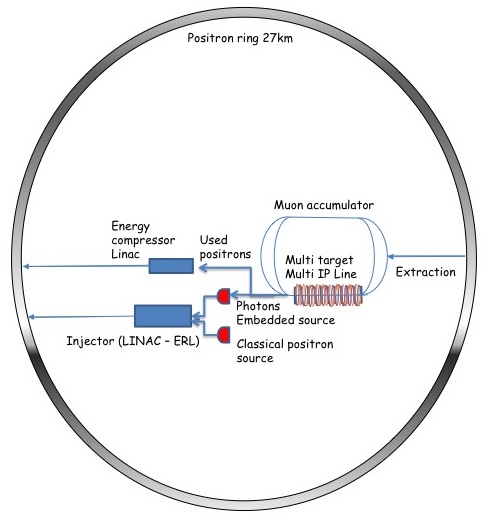}
    \caption{Layout of Scheme I.}
    \label{fig:Scheme1}
\end{figure}

The $e^+$ injection is assured by a high energy superconducting Linac or an Energy Recovery Linac (ERL). Part of it can be eventually also used as a compressor Linac, depending on the final scheme configuration. 
Fig. \ref{fig:Scheme1} shows the layout of the complex for Scheme I.

\subsection{Timeline of Scheme I}
\begin{figure*}[htpb]
    \centering
    \includegraphics*[width=\textwidth]{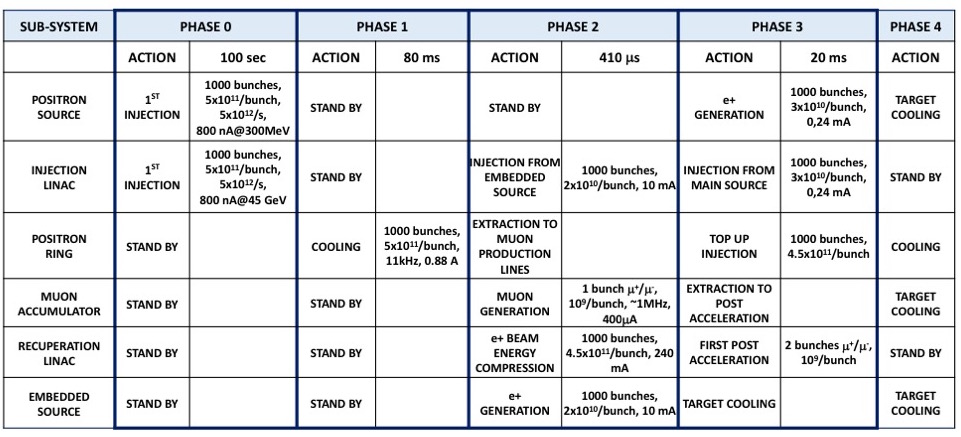}
    \caption{Timeline of Scheme I.}
    \label{fig:Timeline1}
\end{figure*}

The timeline of the different scheme can be illustrated by dividing the full muon production cycle in different phases, each identified by the action of each of the systems and by the duration. To better describe the full muon production the first phase takes also into account the first production cycle, when the main positron ring is filled with the first train of 1000 positron bunches. In Fig.\ref{fig:Timeline1} all the phases described below are illustrated. \par
All 3 schemes will have a common Phase 0, when the positrons are produced and accelerated in the Linac, ready to be injected either in a Damping Ring or the main Positron Ring. This process is supposed to take up to 100 sec.

\begin{Itemize}
\item	Phase 0: during this phase the $e^+$ source and the injection Linac have to produce, accelerate to 45 GeV, and inject 1000 bunches of $5x10^{11}$ $e^+$, corresponding to a $5x10^{12}$ $e^+$/s source (SLC $e^+$ source performance in the '90) and an 800 nA average current beam. This can be fulfilled in a long period to reduce the average current constraints on the $e^+$ sources and the injection Linac. The cycle duration has arbitrarily been fixed at 100 sec (it can be shortened). All the other systems are in stand-by mode in this phase.
\item	Phase 1: begins with the cooling of the beam in the main Positron Ring for two damping times, corresponding to 80 ms. This time can be shortened if a low emittance $e^+$ source is designed coupled with the adiabatic damping emittance reduction. In this phase the 27 km – 11 kHz period ring is filled with a 0.88 A average current beam with $\sim$ 100 MW synchrotron power. All the other systems are in stand by.
\item	Phase 2: the muon production happens in 410 $\mu$A. All the 1000 ``fresh'' $e^+$ bunches are extracted with a delay of 410 ns (delay loops will allow for synchronization with the muon bunches) and injected in the targets line. Each passage will produce $\mu^{+}$ and $\mu^{-}$ bunches that are re-circulated in two Muon Accumulator (MA) rings, $\sim$120 m long, synchronously with the next $e^+$ bunches. The high energy gamma rays flux also coming from the $e^+$ collision on the targets can be used to produce $e^+$ in an ``embedded source''. After the target interactions the $e^+$ bunches are ``chirped'' in a magnetic chicane (i.e. a correlation is added to the longitudinal phase space distribution of the $e^+$ beam) and then pass through the energy compressor Linac, to be injected back in the $e^+$ PR, so minimizing the beam losses. Taking into account a $\sim$90\% efficiency in these $e^+$ bunches recuperation, the compressor Linac has 240 mA of pulsed current and 720 $\mu$A average at 10 Hz. At the same time the embedded $e^+$ source should provide $2.5x10^{10}$ $e^+$ (depending on its efficiency) each, for a pulsed current of 13 mA in the injection Linac. The muon rings are accumulating muons up to a bunch intensity of $10^{9}$ particles that, taking into account a MA ring period of 410 ns gives a current of $\sim$400 $\mu$A. The main $e^+$ source is in stand-by.
\item	Phase 3: the main $e^+$ source produces and provides to the injection Linac the missing 1000 $e^+$ bunches with $2.5x10^{10}$ particles (depending on the embedded source efficiency) that are injected in top in the $e^+$ ring in $\sim$20 ms. The muon bunches are extracted and sent to the post acceleration cycle. The PR is filled up to the nominal bunch
charge of $5x10^{11}$ $e^+$ per bunch. In the embedded source the target cools down.
\item	Phase 4: end of the cycle, restarts from Phase 1. In this phase the main and embedded $e^+$ source targets and muon targets will be cooling down. 
\end{Itemize}

\section{Scheme II }
\subsection{General layout}
\begin{figure}[htpb]
    \centering
	\includegraphics[width=\columnwidth]{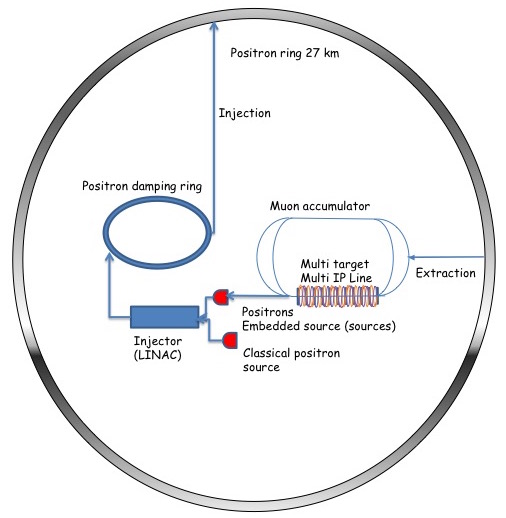}
    \caption{Layout of Scheme II.}
    \label{fig:Scheme2}
\end{figure}

The second scheme aims to take into account some of the critical points and parameters of the first one. Namely, the possibility to ramp up and down the PR magnets allows to drastically reduce the emitted synchrotron power and to substitute the 45 GeV injector Linac with a system composed by a 5 GeV injector Linac and a 5 GeV, 27 km long, Damping Ring (DR) in the same PR tunnel. The $e^+$ will be always produced by both the classical source and the embedded one. Nevertheless, in this scheme the last one does not use, as drive beam, the emitted gammas from the muon target line but the ``spent'' $e^+$  bunches coming from the production targets. The PR and the muon accumulation rings remain basically the same of the previous scheme. The main $e^+$ source is mandatory to produce the first generation of the 1000 positron bunches in the Damping Ring. Fig.\ref{fig:Scheme2} shows the layout of the complex for Scheme II. 
\subsection{Timeline of Scheme II}
\begin{figure*}[htpb]
    \centering
    \includegraphics*[width=\textwidth]{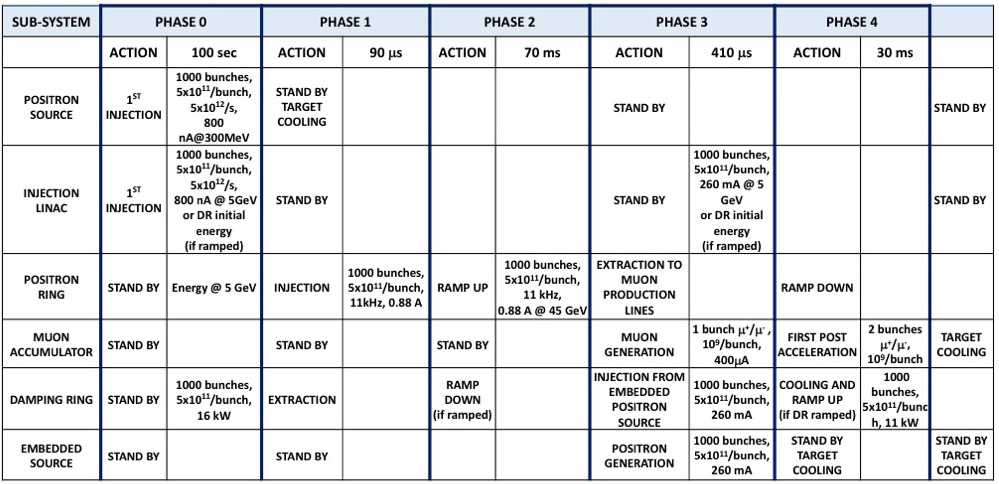}
    \caption{Timeline of Scheme II.}
    \label{fig:Timeline2}
\end{figure*}

The timeline of the second scheme can be divided in five different phases, illustrated in Fig.\ref{fig:Timeline2}. This Scheme will start with the same Phase 0 than Scheme I (not shown in Fig.\ref{fig:Timeline2}).
\begin{Itemize}
\item	Phase 0: $e^+$ production and acceleration to 5 GeV same as in Scheme I, plus the injection and cooling of 1000 bunches of $5x10^{11}$ $e^+$ in the Damping Ring. This stores 3.8 A positron current and provides a short cooling time (30 msec) thanks to damping wigglers. All the other systems are in stand-by mode in this phase.
\item	Phase 1: the positron beam is extracted in 90 $\mu$s from the DR and injected in the main PR that has been previously ramped down to the DR energy. The circulating current in the PR is always 0.88 A, but the synchrotron power is reduced by a factor of $\sim$6500. All the other systems are in stand-by.
\item	Phase 2: in this phase the PR energy is ramped up to 45 GeV in 70 ms. At the end of this phase the PR is ready for the muon production. Should also the DR be working in a ramping up and down cycle, this phase will provide the time to ramp the DR down to the positron injection energy from the positron source and Linac (5 GeV). All the other systems are in stand-by. 
\item	Phase 3: the muon production happens, as before, in 410 $\mu$s. As far as the muon production is concerned the scheme and the systems parameters are the same of the ones illustrated in phase 3 of scheme I. In this phase, differently from scheme I, after the target interactions, the ``spent'' $e^+$ bunches are sent to the main positron source to provide the regeneration of all the positron bunch population, having thus a 100\% efficiency ($e^+$ on target/$e^+$ captured). At 10 Hz the positron source is supposed to provide $10^{16}$ $e^+$/sec. The pulse current in the injector is 260 mA. At the end of this phase the DR is filled again with 1000 bunches at the nominal intensity.
\item	Phase 4: in 30 msec the DR provides for the $e^+$ beam cooling. This time must also take into account the possibility of ramping its energy up and down. In parallel the PR energy is ramped down to 5 GeV, ready for the injection from the DR. The positron source targets will cool down for 100 msec, corresponding to this phase and the 70 ms of Phase 3. After this, the cycle is repeated again from Phase 1. 
\end{Itemize}
\section{Scheme III }
\begin{figure*}[ht]
    \centering
    \includegraphics*[width=\textwidth]{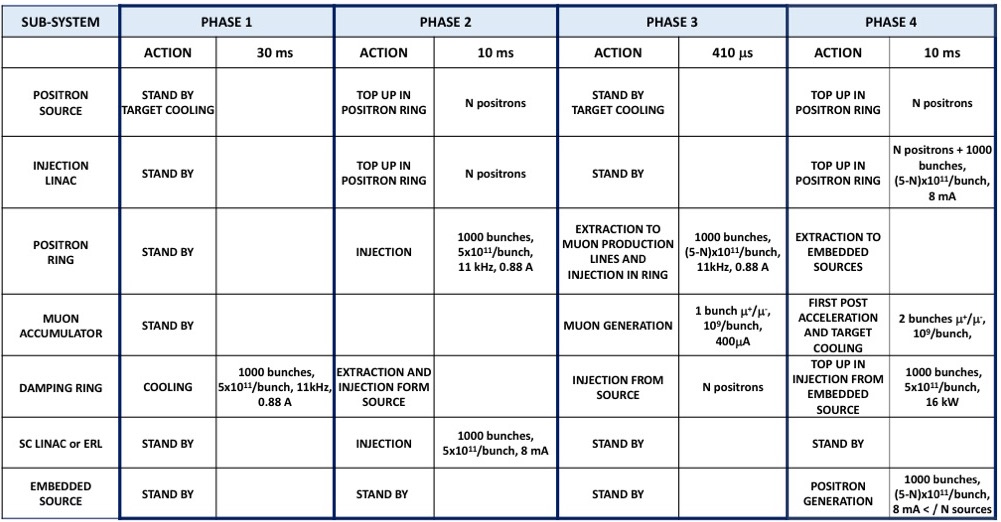}
    \caption{Timeline of Scheme III. Phase 0, same as in Scheme I, is not shown here.}
    \label{fig:Timeline3}
\end{figure*}

\subsection{General layout}
\begin{figure}[htpb]
    \centering
	\includegraphics[width=\columnwidth]{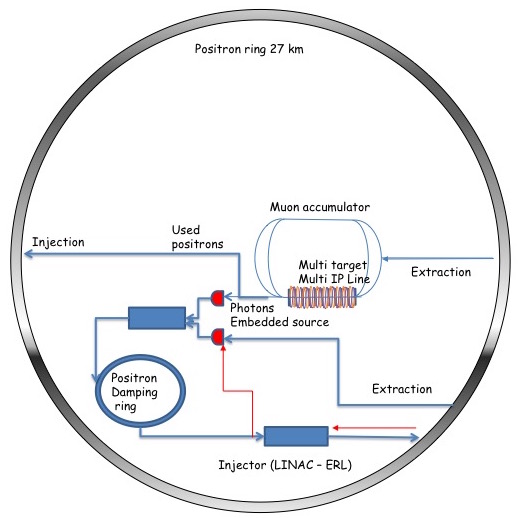}
    \caption{Layout of Scheme III.}
    \label{fig:Scheme3}
\end{figure}

The third scheme intends to take into account some of the critical points and parameters of the first two. It is always envisaged the use of a Damping Ring to cool down the $e^+$ produced from the $e^+$ source. In scheme III, a design for a partial recuperation of the used $e^+$ is included, with an estimated efficiency of 70\% by injecting the spent $e^+$ bunches directly back to the positron storage ring. This allows to have a slow extraction (20 msec) of the $e^+$ for the following $e^+$ production, so avoiding to operate the Linac systems in 410 $\mu$sec with a consequent extremely high value of the pulse current. The injection in the main ring is provided, as in scheme I, by a high energy SC Linac or ERL. Like in scheme II, the lost $e^+$ are replaced by using the ``spent'' $e^+$ beam after the muon generation and in parallel to the main $e^+$ source. Since the efficiency of the capture system of the $e^+$ source is increased in the case of a drive beam at a lower energy with respect to 45 GeV, the possibility of a deceleration phase in the injection Linac is also taken into account. No beam will be circulating in the PR during the DR cooling phase, so the synchrotron radiation emission duty cycle is reduced, decreasing also the synchrotron power budget with respect to scheme I. The repetition rate of one full cycle is 20 Hz. Fig.\ref{fig:Scheme3} shows the layout of the complex for Scheme III. 
\subsection{Timeline of Scheme III}
The timeline of the third scheme can be divided in five different phases, illustrated in Fig.\ref{fig:Timeline3}. This Scheme will start with the same Phase 0 as in Scheme I (not shown in Fig.\ref{fig:Timeline3}).
\begin{Itemize}
\item	Phase 0: $e^+$ production same as in Scheme I.
\item	Phase 1: the 1000 bunches of $5x10^{11}$ $e^+$ are cooled in 30 ms in the DR.
\item	Phase 2: after the cooling the $e^+$ beam is extracted from the DR and injected in the main PR in 10 ms, thus reducing the pulsed current of the injector to 8 mA. All the other systems are in stand-by, except the main positron source which can restart to produce and accelerate $e^+$ to re-inject in the DR, to partially restore the $e^+$ beam losses.
\item	Phase 3: the muon production happens, as before in 410 $\mu$sec. As far as the muon production is concerned the scheme and the systems parameters are the same of the ones illustrated in phase 3 of scheme I and II. In this phase, differently from scheme I and II, after the target interactions, the ``spent'' $e^+$ bunches will be sent back to the main PR, with a reduced injection efficiency due to the high energy spread generated in the targets interactions. It is estimated that the PR momentum acceptance will allow for $\sim$70\% of this ``spent'' beam to be stored. At the end of this phase the DR is filled again with 1000 bunches at reduced intensity.
\item	Phase 4: the $e^+$ bunches are slowly extracted from the PR, decelerated in the injector Linac, and sent back to the DR
to be cooled. In the meantime the conventional Positron Source will restore the missing $e^+$ in the DR (topping up). In this phase the produced muons bunches are extracted and post accelerated. Also in this phase the Linac system has a 8 mA pulse current intensity. After this phase the cycle is repeated from Phase 1, while the $e^+$ main and embedded source targets and the muon targets cool down and all the other system are in stand-by. 
\end{Itemize}

\section{ACCELERATOR COMPLEX SUB-SYSTEMS}
Once the three schemes are developed, an evaluation of the possible performances of all the main sub-systems is mandatory in order to asses the consistencies between the declared parameters and their actual feasibility or to identify the possible future R\&D program.
In the following we will detail the characteristics of the different sub-systems of the accelerator complex, for the 3 Schemes considered.

\begin{figure*}[!h]
    \centering
    \includegraphics*[width=0.7\textwidth]{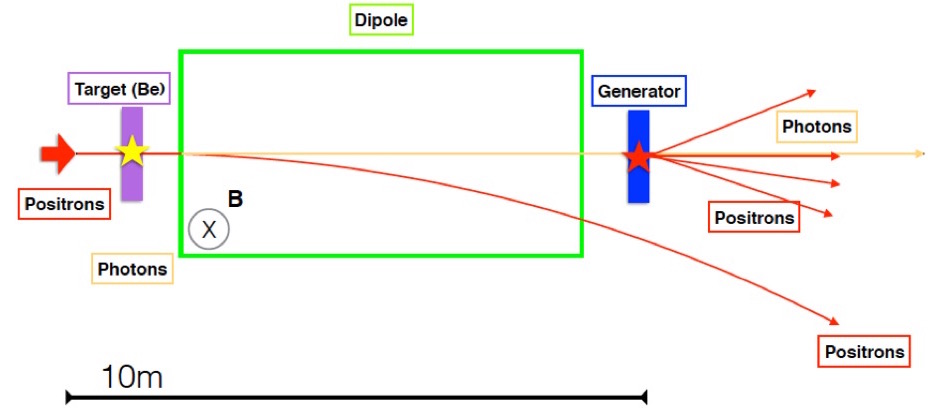}
    \caption{Embedded source scheme.}
    \label{fig:Fig16}
\end{figure*}

\section{POSITRON SOURCES}
The $e^+$ source has to provide trains of 1000 bunches with $5x10^{11}$ $e^+$/bunch to inject in the Damping Ring at 5 GeV. For the initial injection there are no time constraints, assuming an $e^+$ source like the ILC \cite{ILC} or CLIC one \cite{CLIC}, which are designed to produce $10^{14}$ $e^+$/sec, the injection will take 5 seconds.
On the contrary, the source needed to replace the $e^+$ lost in the muon production process is a real challenge, since the time available to produce, damp and accelerate the $e^+$ is very short.
If we consider for example Scheme III, we can assume that 70\% of the $e^+$ at the targets exit can be recovered, injected in the main $e^+$ ring, slowly extracted and decelerated and injected in the Damping Ring. 

 Therefore only 30\% of the required $e^+$ need to be produced by the source in a time cycle t$_{cycle}$ = 50 msec, corresponding to the 20 Hz repetition frequency. We assume to inject the bunches in the damping ring during 20 msec and to store them for 30 msec (three damping times) to damp the emittance. The required $e^+$ production rate is then $3x10^{15}$ $e^+$/sec. In order to achieve such a high rate of $e^+$ production we need to explore all the techniques developed for the future linear colliders like hybrid targets (crystal target + tungsten target) \cite{CLIC} and rotating targets \cite{ILC} and we will develop an R\&D program on new targets. 
The DR energy acceptance has to be very large, at least as large as the main $e^+$ ring ($\pm$6\%). The present lattice satisfies this requirement.
An optimization of the $e^+$ capture system in order to take advantage of the large DR energy acceptance could improve the $e^+$ yield.
Another possibility to reduce the requested $e^+$ rate is to increase the energy acceptance of the PR to reduce the fraction of lost $e^+$ to be replaced by the source.
If all these efforts would not succeed in providing the required $e^+$ rate we could use more than one source.
\begin{figure}[h]
    \centering
	\includegraphics[width=0.9\columnwidth]{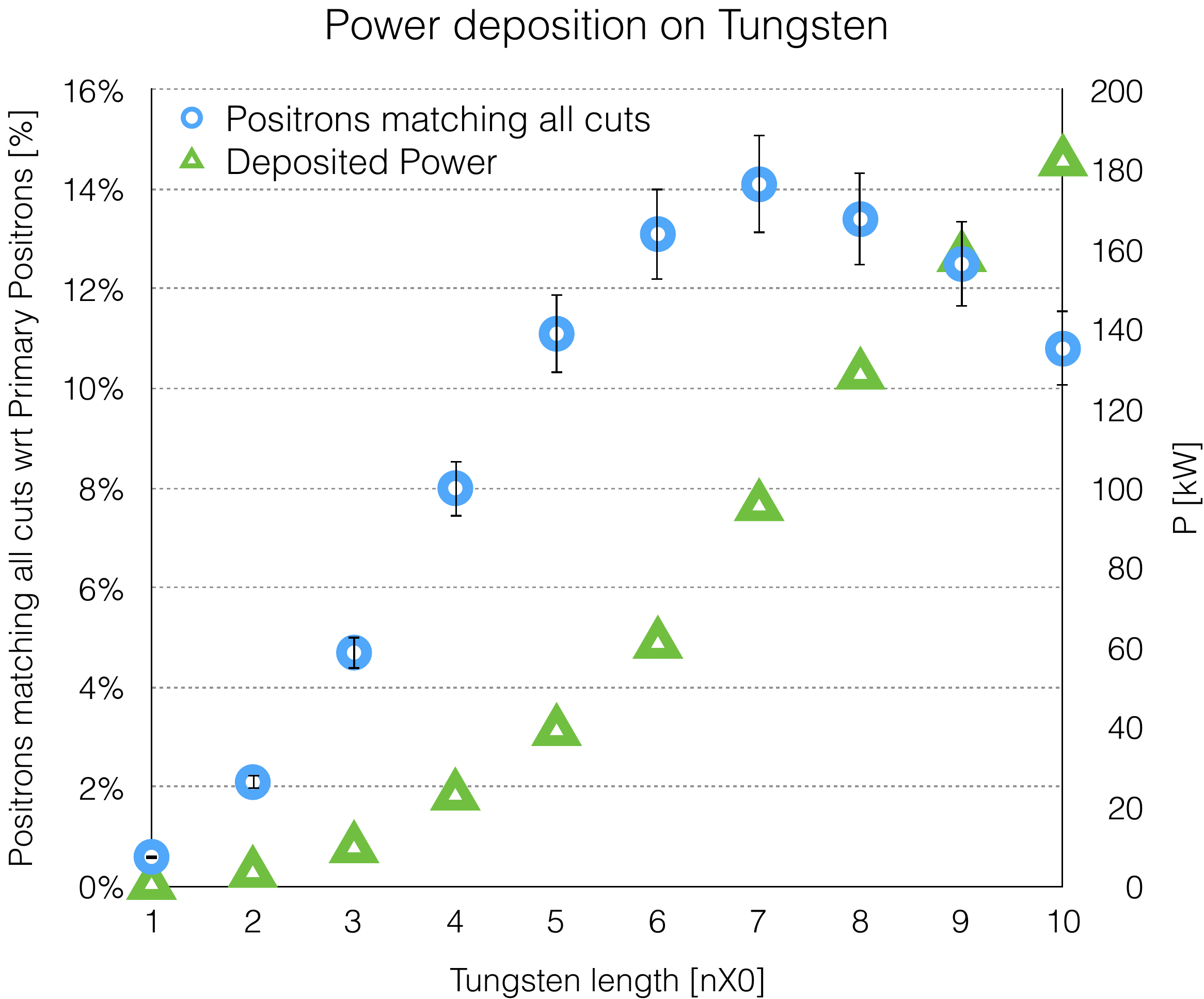}
    \caption{Fraction of $e^+$ matching all quality cuts (see text) and power deposited in the Tungsten target as a function of its thickness.}
    \label{fig:ps2}
\end{figure}
\begin{figure*}[h]
    \centering
    \includegraphics*[width=0.8\textwidth]{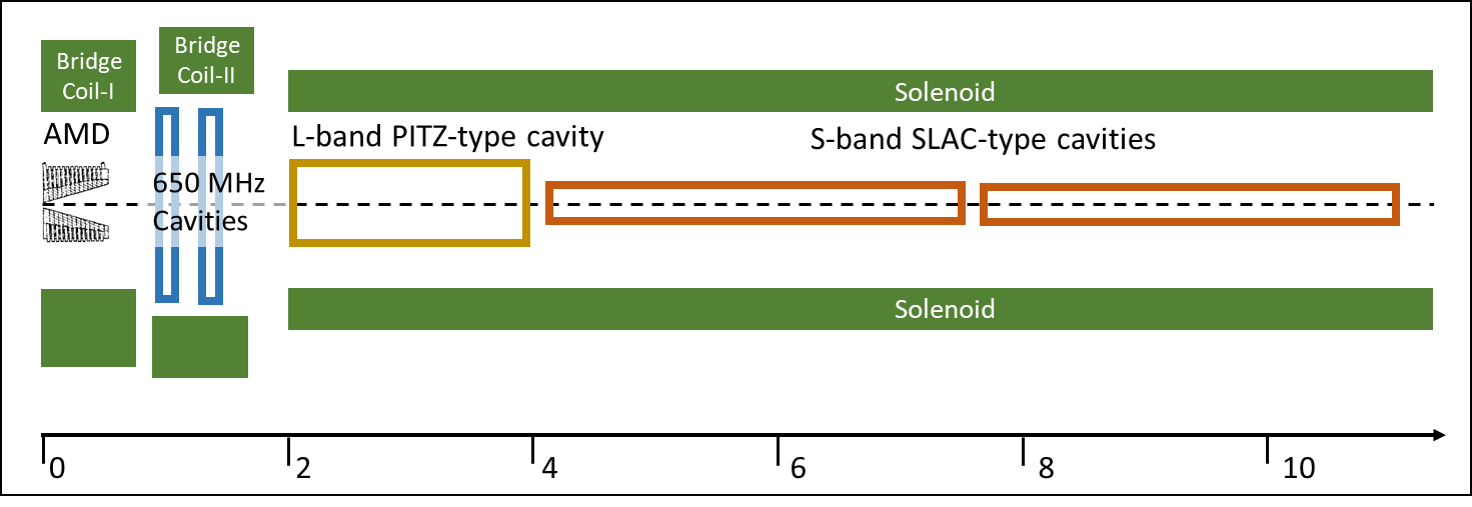}
    \caption{Layout of the capture system.}
    \label{fig:ps3}
\end{figure*}

\subsection{Embedded Source}
The 45 GeV $e^+$ passing through the muon targets produce a large number of high energy photons. We have studied the feasibility of an embedded source \cite{Collamati} that uses these photons exiting after the dipole, which bends away the $e^+$, to produce new $e^+$ impinging on a 5 radiation length (5X$_0$) tungsten target (see Fig.\ref{fig:Fig16}). 
For each positron on the primary 3 mm Be target there are:
\begin{Itemize}
\item	0.11 photons hitting the W target
\item   0.65 $e^+$ coming out of the 5X$_0$ W target.
\end{Itemize}
To give a preliminary estimate of the number of collected $e^+$, the number of $e^+$ within the parameter range that could be typically accepted by the capture system (i.e. energy between 5 MeV and 20 MeV, positions < 0.5 cm, angles < 0.5 rad) has been evaluated as a function of the target thickness (see Fig.\ref{fig:ps2}). 

For 5X$_0$ target thickness about 11\% of the $e^+$ match all the cuts, corresponding to a yield of 0.07. The power deposited on the target for 1000 bunches with $5x10^{11}$ $e^+$/bunch is 39 kW at the 10 Hz repetition rate. 
The power deposited on the target as a function of the target thickness is also shown in Fig.\ref{fig:ps2}, highlighting how the achievable positron yield depends on the power that the target can sustain. In order to give a more precise estimate of the achievable yield, a simulation of the positron capture system has been performed. 

\subsubsection{Simulation of the positron collection efficiency}
A simulation of the collection efficiency has been performed with ASTRA tracking code \cite{ASTRA},
representing the $e^+$ bunch, at the target source, with $20\:k$ macro-particles. 

The layout of the capture system is shown in Fig.\ref{fig:ps3}: the tungsten target is followed by an Adiabatic Matching Device (AMD) \cite{chehab1994}, a pulsed solenoidal lens with a longitudinal variation of the magnetic field given by $B(z)=B_0/(1+\mu z)$, with  a peak field $B_0$ = $8\: T$ and $\mu=50\: m^{-1}$, for a length of 30 cm.
Downstream the AMD there are two single cell $650\:MHz$ cavities, followed by a 14 cells L-band accelerating cavity, PITZ-like \cite{Paramonov:LINAC10-MOP081}. 
Then two S-band SLAC type cavities are used to accelerate the beam to $190\: MeV$, ready for the last boosting stage.
The AMD and all accelerating cavities are immersed in a static magnetic field of $0.5\: T$. 
The beam pipe aperture and the L-band cavity have a radius $r=20\: mm$, while inside the S-band cavities $r=10\: mm$. The whole capture channel, schematically shown in Fig.\ref{fig:ps3}, is $\sim$10 meters long. In the simulation all the cavities work on crest.

\begin{figure}[!h]
    \centering
	\includegraphics[width=0.8\columnwidth]{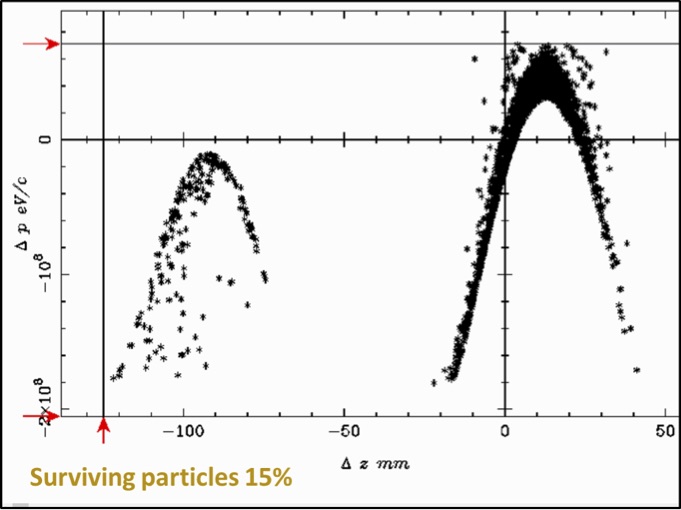}
    \caption{Longitudinal distribution of surviving particles: 2926 mp on 20k simulated. The most populated rf bucket brings about $95\%$ of all the surviving particles.}
    \label{fig:ps5}
\end{figure}

In a preliminary optimization 15\% of the $e^+$ are collected for 5X$_0$ target thickness, corresponding to a yield of 0.1. The longitudinal distribution of the surviving particles is shown in Fig.\ref{fig:ps5}, while the main beam parameters are reported in Tab.\ref{tab:capture_out}

Next step is to perform an optimization of the AMD, capture section and first linac sections using the genetic algorithm code GIOTTO \cite{GIOTTO}.
Due to the high photon energy a thick target is required to increase the positron production in this configuration: the main challenge is the high power deposition in the target.

\begin{table}[h!]
	\begin{center}
		\caption{Beam parameters at the end of the embedded source capture channel. The bunch length $\sigma_{z}$ refers only to particles trapped in the most populated bucket, as shown in Fig. \ref{fig:ps5}}
		\label{tab:capture_out}
		\begin{tabular}{|l|l|}
		    \hline
			\textbf{Parameter}  & \textbf{Value}            \\ \hline
			survived macroparticles & $2926\: (on\: 20k)$        \\ \hline
			$\sigma_{z}$            & $9.6 \ mm\ (most\ pop.\ bucket)$              \\ \hline
			$\sigma_{x}$            & $4.0\ mm$              \\ \hline
			$\sigma_{y}$            & $3.9\ mm$              \\ \hline
			$\varepsilon_{n,x}$     & $8.2\times 10^3$ mm mrad            \\ \hline
			$\varepsilon_{n,y}$     & $7.9\times 10^3$ mm mrad            \\ \hline
			$E$                & $\simeq 190\ MeV$ \\ \hline
			$\Delta_E$    & $4.9\times 10^{4}\ keV$\\ \hline
		\end{tabular}
	\end{center}
\end{table}

\section{MAIN POSITRON RING}
The PR at 45 GeV should have small beam emittance, mostly round beams, and a large energy acceptance in order to be able to accommodate the “spent” beam coming back after the muon production. Damping time should not be an issue since on-axis injection is foreseen and the ``fresh'' positron beam from the source can be cooled in a Damping Ring (DR). \par
The choice of the final lattice will be based on the one showing the larger energy acceptance, since it is mandatory that possibly all the “spent” beam from the muon production be successfully re-injected in the PR to be later decelerated and re-injected in the DR for cooling.\par
A large circumference can accommodate a large number of bunches with less important synchrotron losses. In order to accommodate the requested 1000 bunches with $5x10^{11}$ $e^+$/bunch a 27 km LHC-like was preferred. However a solution for a 100 km ring FCC-like could increase the muon collider luminosity by a factor of 3 at least.\par
\begin{figure}[h]
    \centering
	\includegraphics[width=0.9\columnwidth]{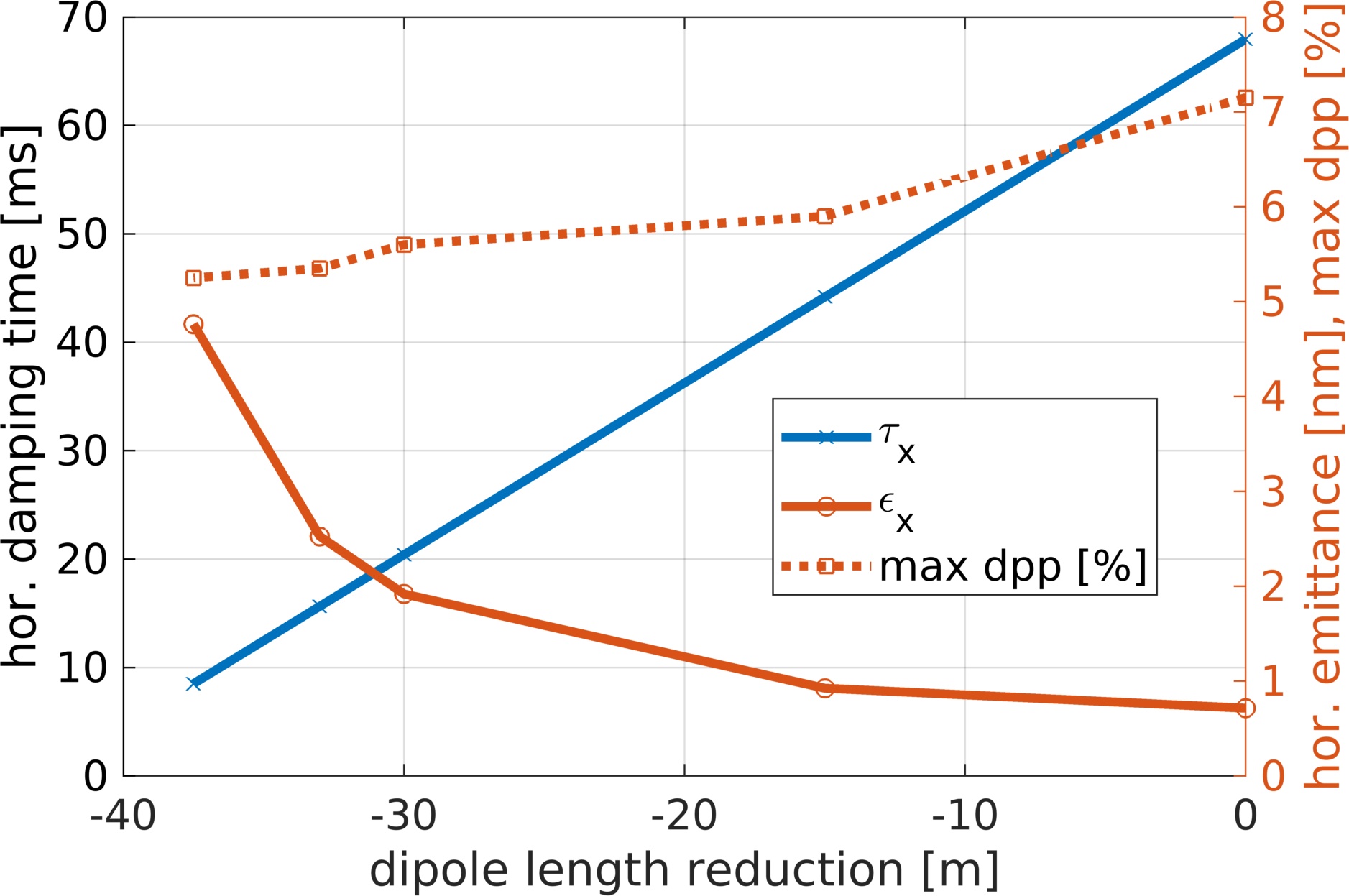}
    \caption{Damping time, emittance and energy acceptance behavior as a function of the total dipole length reduction in a cell for the 0.7 nm, 64 cells lattice.}
    \label{fig:Fig13}
\end{figure}

\begin{figure*}[ht]
    \centering
    \includegraphics*[width=\textwidth]{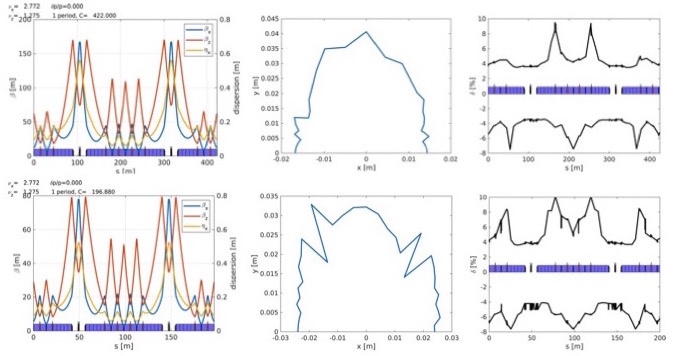}
    \caption{Top: optical functions (left), dynamic aperture (center) and energy acceptance along the ring (right), for a 0.7 nm emittance, 64 cells PR lattice. Bottom: optical functions (left), dynamic aperture (center) and energy acceptance along the ring (right), for a 5 GeV, 70pm emittance Damping Ring. }
    \label{fig:Beta}
\end{figure*}
At present several 27 km long lattices have been studied, with a horizontal emittance ranging from 0.7 to 20 nm. Their energy acceptance ranges from $\pm$2 to $\pm$8 \%, and work is in progress to improve it. The lattices are all inspired to the ESRF upgrade hybrid multi-bend achromat lattice \cite{ESRF}, with different number of cells (64 or 32) and with different dipoles length in order to tune emittance and damping time. This design is based on increasing the number of dipoles/cell, in order to decrease the emittance, long and weaker to decrease the emitted synchrotron radiation. Dipoles in one cell have different lengths and this parameter can be adjusted in order to tune damping time and horizontal emittance to desired values. Fig.\ref{fig:Fig13} shows, for the 64 cells 0.7 nm lattice, the behavior of horizontal damping time, horizontal emittance and energy acceptance as a function of the reduction of the dipoles total length in one cell (844 m long). Table \ref{table:PR} shows a PR reduced parameter list for three lattice designs with different horizontal emittance. As an example, in Fig.\ref{fig:Beta} (top plots) the optical functions, dynamic aperture and energy acceptance along the ring for the 0.7 nm horizontal emittance lattice are shown.

\begin{table}[!h]
   \centering
   \caption{45 GeV PR parameters for 3 different emittances }
   \begin{tabular}{lccc}
       \toprule
       \textbf{Parameter}     & \textbf{0.7 nm}  & \textbf{6 nm} & \textbf{10 nm} \\
       \midrule
          Circumference [km]     & 27            & 27            & 27          \\ 
          N. cells               & 64            & 32            & 32          \\ 
          I$_b$ [A]              & 0.89          & 0.89          & 0.89       \\          
          N$_{part}$ /bunch      & $5x10^{11}$   & $5x10^{11}$   & $5x10^{11}$ \\ 
          N. bunches             & 1000          & 1000          & 1000        \\ 
          E$_{loss}$/turn [GeV]  & 0.12          & 0.12          & 0.19        \\ 
          Nat. $\sigma{_z}$ [mm] & 1.9           & 3.6           & 3.8         \\ 
          $\alpha{_c}$           & 2.9x10$^{-5}$ & 1x10$^{-4}$   & 1.1x10$^-4$ \\  
          Energy spread          & 7x10$^{-4}$   & 7x10$^{-4}$   & 9x10$^{-4}$ \\          
          $\tau_{x,y}$ [ms]      & 68            & 66            & 42          \\  
          Energy acceptance [\%] & $\pm$ 8       & $\pm$6      & $\pm$ 2   \\ 
          SR power [MW]          & 106           & 109           & 170         \\  
\bottomrule
   \end{tabular}
   \label{table:PR}
\end{table}

\section{DAMPING RING}
The introduction of a DR into the muon source chain is advisable, so relaxing the requests on the PR. 
The DR should provide fast cooling of the $e^+$ produced by the source. A 5 GeV, 6.3 km DR, could provide the requested damping time (about 10 msec at 5 GeV) and large energy acceptance, with a number of damping wigglers to increase the energy losses.  
As an example, a preliminary design was done based on the same lattice as the PR, with 32 cells and an emittance of 70 pm. Table \ref{table:DR} shows the DR parameters, assuming the insertion of about 100 damping wigglers similar to those in the ILC TDR \cite{ILC}.
\begin{table}[!h]
   \centering
   \caption{5 GeV DR parameters with ILC-like wigglers }
   \begin{tabular}{lc}
       \toprule
       \textbf{Parameter}     & \textbf{70 pm}   \\
       \midrule
          Circumference [km]                   & 6.3                  \\ 
          N. cells                             & 32                   \\ 
          I$_b$ [A]                            & 3.8                \\ 
          Horizontal emittance [pm]            & 70                   \\
          Coupling factor                      & 0.5                  \\
          N$_{part}$ /bunch                    & $5x10^{11}$          \\ 
          N. bunches                           & 1000                 \\ 
          E$_{loss}$/turn with wigglers [GeV]  & 1.8x10$^{-2}$         \\ 
          Nat. $\sigma{_z}$ [mm]               & 6                    \\ 
          $\alpha{_c}$                         & 1.2x10$^{-4}$        \\  
          E spread                             & 3x10$^{-4}$          \\          
          $\tau_{x,y}$ with wigglers [ms]      & 10                   \\  
          Energy acceptance [\%]               & $\pm$ 10             \\ 
          SR power with wigglers [MW]          & 67                   \\  
\bottomrule
   \end{tabular}
   \label{table:DR}
\end{table}
In Fig.\ref{fig:Beta} (bottom plots) the optical functions, dynamic aperture and energy acceptance along the ring for the 70 pm emittance, 5 GeV DR lattice are shown. The energy acceptance for this very preliminary lattice is very promising, spanning from a minimum of $\pm4$$\%$ to a maximum of $\pm10$$\%$.

 In order to reduce the number of damping wigglers needed a shorter ring is preferable, however in case the option to use the PR as a DR is chosen, by ramping down its energy, an example DR with 27 km and 9 GeV has also been studied. In this case the PR lattice (with 64 cells, emittance 5.8 nm and damping time 4 msec at 45 GeV) the sign of the magnetic field in 2 of the 5 dipoles in a cell will have to be inverted during the ramping. With the addition of about 500 damping wigglers, a damping time of 11 msec can be achieved at 9 GeV, with an emittance of 2.7 nm. In case this option is chosen, a careful study of the ramping procedure of the dipole-quadrupole magnets should be performed.
 
\section{COMPRESSOR LINAC}
\begin{figure}
   \centering
    \includegraphics[width=\columnwidth]{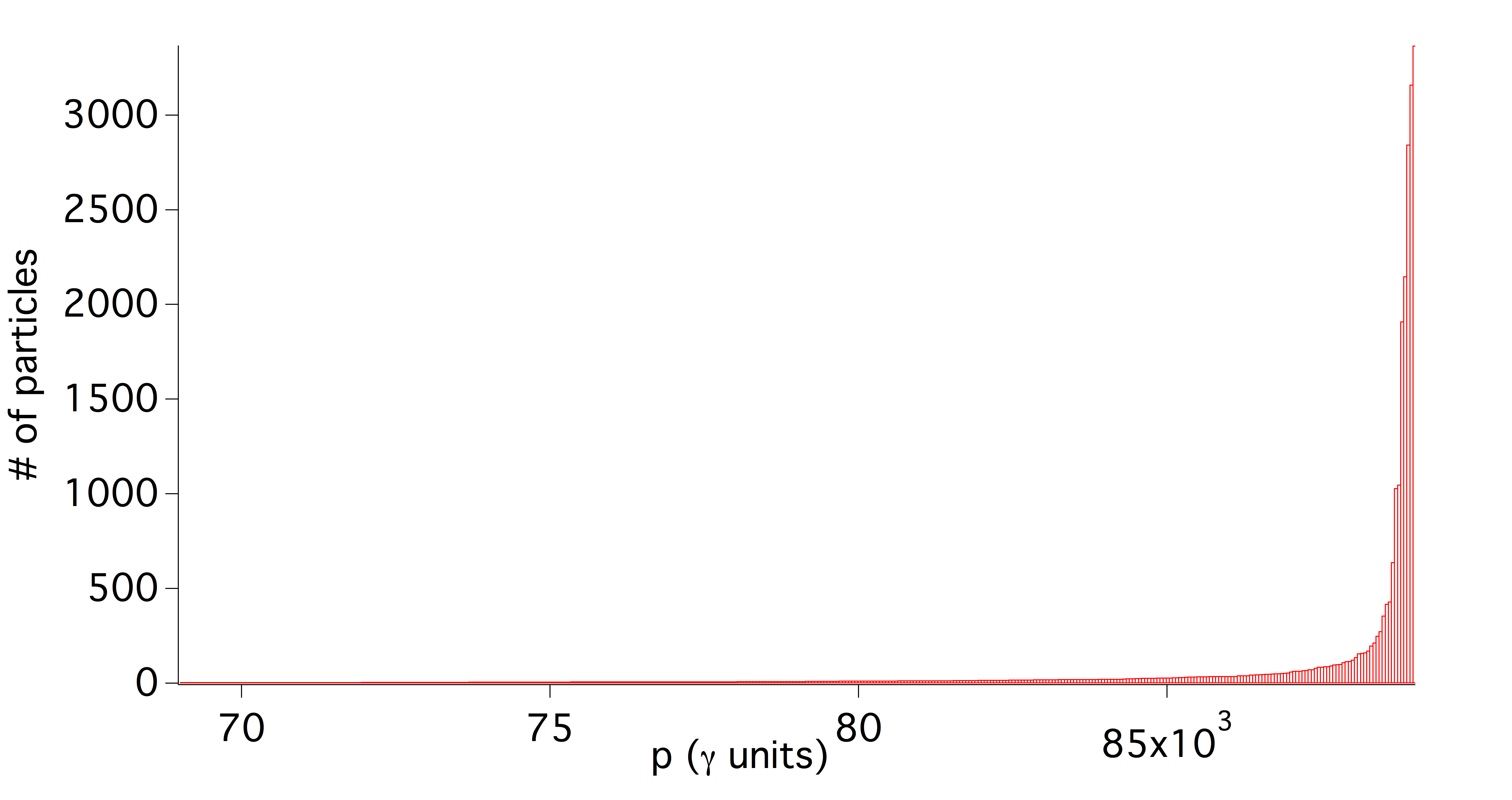}
    \caption{Energy distribution of a 90\% of a 30000 particles beam after the interaction with 10 W targets, 3 mm thick, as resulting from Geant simulation.  }
    \label{fig:Positron_energy_distribution}
\end{figure}

The energy compressor Linac is meant to recover the uncorrelated energy spread of the positron beam after the muon production. The considered case refers to a 45 GeV positron beam after the interaction with ten W targets, 3mm thick. In Fig. \ref{fig:Positron_energy_distribution} the energy distribution of the 90\% of the  positron beam is plotted where a maximum energy deviation of $\approx 20\% $ is shown and has been taken as a baseline parameter for the Linac working point.

\begin{figure}
\centering
    \includegraphics[width=\columnwidth]{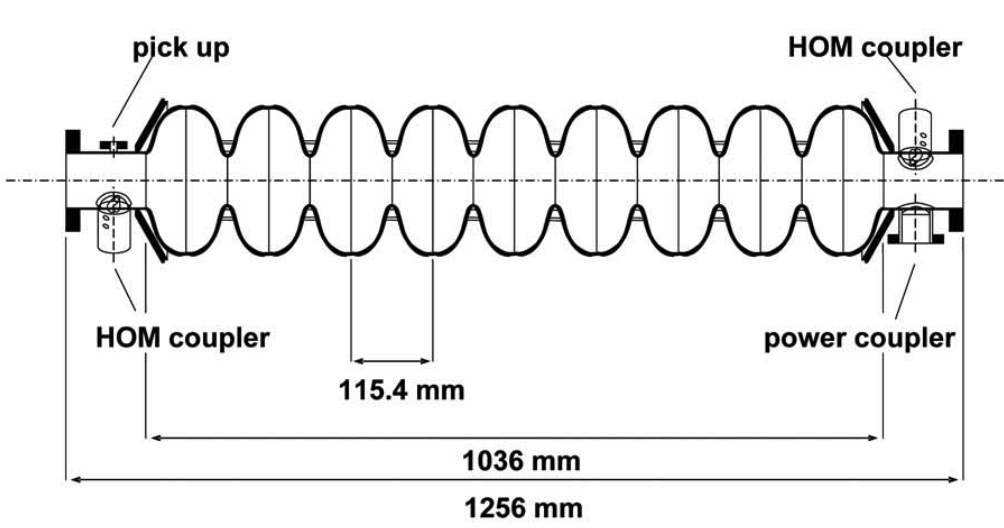}
    \caption{Side view of the nine-cell cavity with the main power coupler port (right), the
pick up probe (left), and two HOM couplers as from \cite{XFEL_TDR}.}
    \label{fig:XFEL_module}
\end{figure}

For a preliminary design of the Linac lattice an L-band structure has been considered for the accelerating module based on the XFEL design as reported in \cite{XFEL_TDR} and shown in Fig.\ref{fig:XFEL_module}, with an average accelerating field of $ E_{acc} \approx 30 MV/m$, while for the longitudinal and transverse short range wakefields the pill-box approximation has been used following the treatment reported in ref. \cite{Kbane}  and  \cite{Zagorodnov}. The Linac layout schematically consists in a $\approx 50 m$ long magnetic chicane upstream the booster to lengthen the beam and provide an energy-position correlation to the positron longitudinal phase space, followed by a 500 m booster operated at the Rf phase $\phi_{RF} =-90^{0} $ for the first 400 m and at $\phi_{RF} =-60^{0} $ for the rest of the linac to recover the average energy at $E_{av}\approx 45 GeV$ and reach a final energy spread of  $\sigma_{\delta} \approx 2 \%$ rms.
\begin{figure}
   \centering
    \includegraphics[width=\columnwidth]{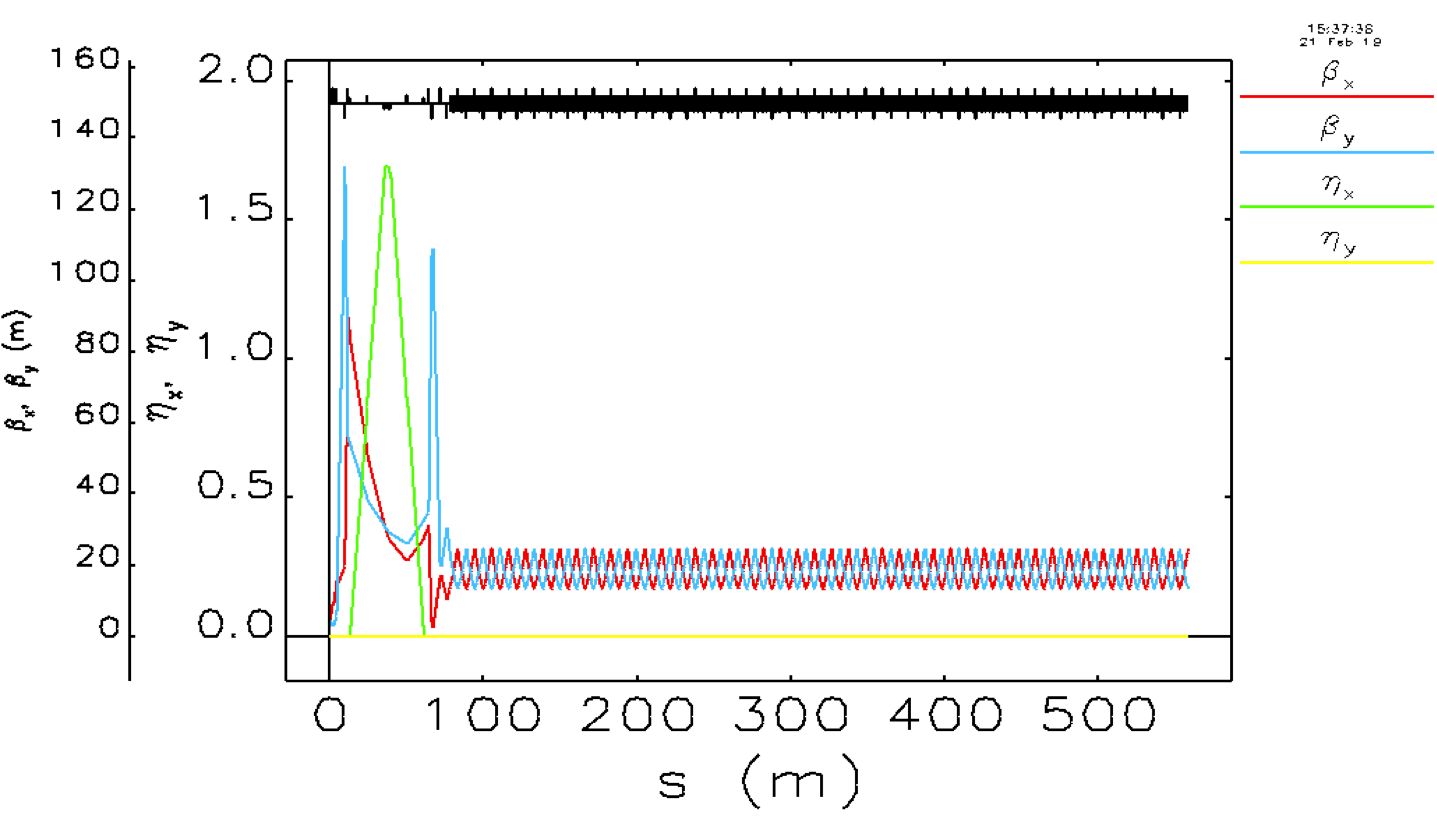}
    \caption{Twiss parameter evolution along the compressor linac. }
    \label{fig:linac_lattice}
    \end{figure}
\begin{figure}
   \centering
    \includegraphics[width=\columnwidth]{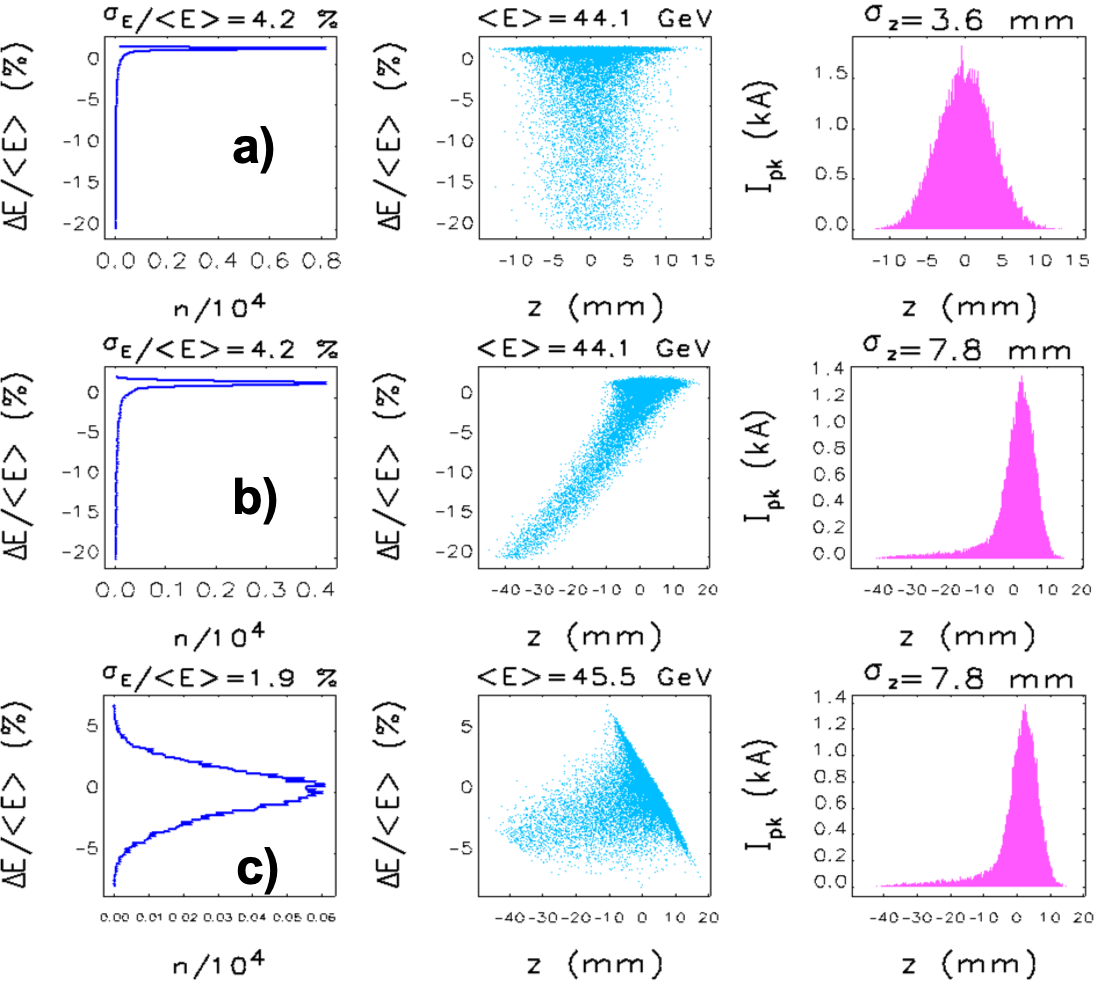}
    \caption{Longitudinal phase space distribution of the positron beam in the compressor linac: a) at the entrance of the matching section upstream the magnetic chicane, b) at the exit of the chicane, c) at the exit of the linac. }
    \label{fig:long_pspace}
\end{figure}

 In Fig. \ref{fig:linac_lattice} the Twiss parameter evolution along the Linac is reported, in this case for the chicane we have $R_{56} = 134 mm$ and $T_{566} = 200 mm$. The longitudinal phase space evolution is reported in Fig. \ref{fig:long_pspace} as resulting from the tracking of 30kp beam with the Elegant code \cite{Elegant}, the longitudinal distribution of the positron beam is shown at the entrance of the first matching section before the chicane a), at the exit of the magnetic chicane b) and finally at  the exit of the linac c). No other considerations have been taken into account up to now except the effects of the chicane $R_{56} $ parameter and the proper phasing of the accelerating field on the longitudinal phase space of the positron beam, nevertheless a more manageable energy spread is achieved at least for the 90\% of the positron beam at the end of the Linac.

\par
\section{TARGET STUDIES}
\begin{figure}[!h]
    \centering
	\includegraphics[width=\columnwidth]{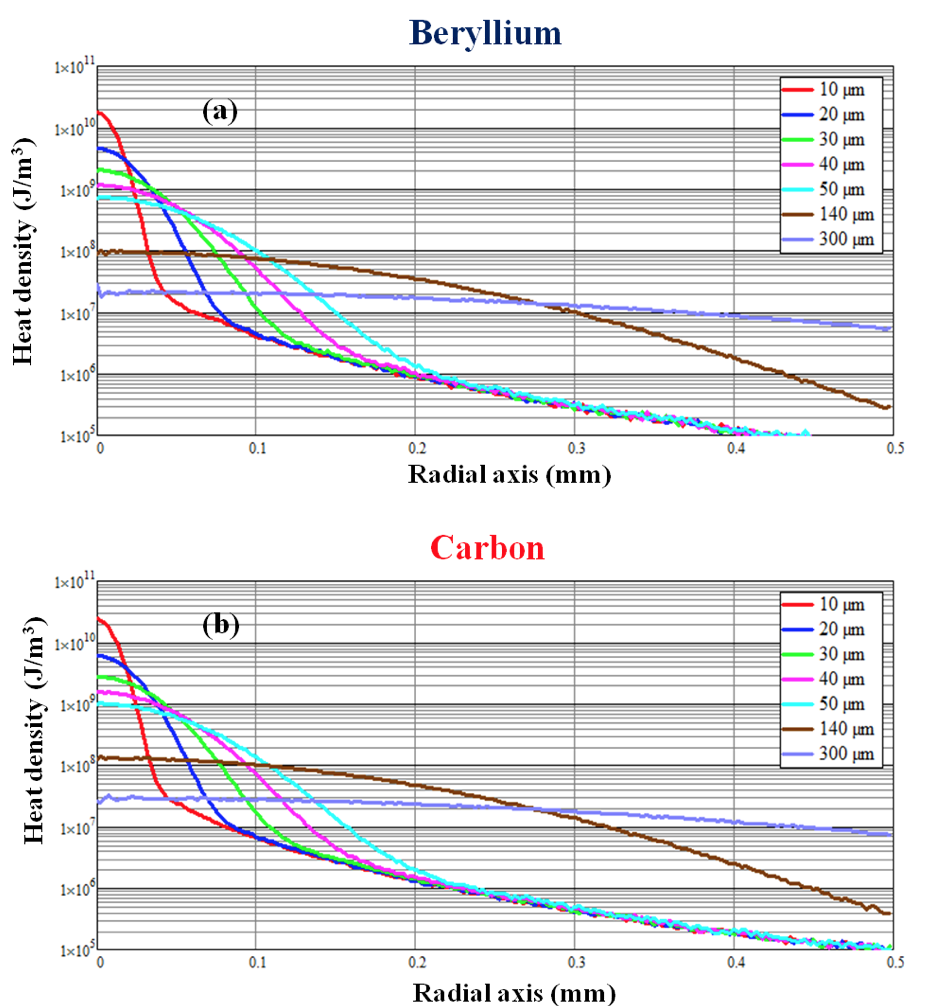}
    \caption{Deposited Energy density by 1 bunch for Beryllium (a) and Carbon (b).}
    \label{fig:pedd}
\end{figure}

\begin{figure}[!h]
    \centering
	\includegraphics[width=\columnwidth]{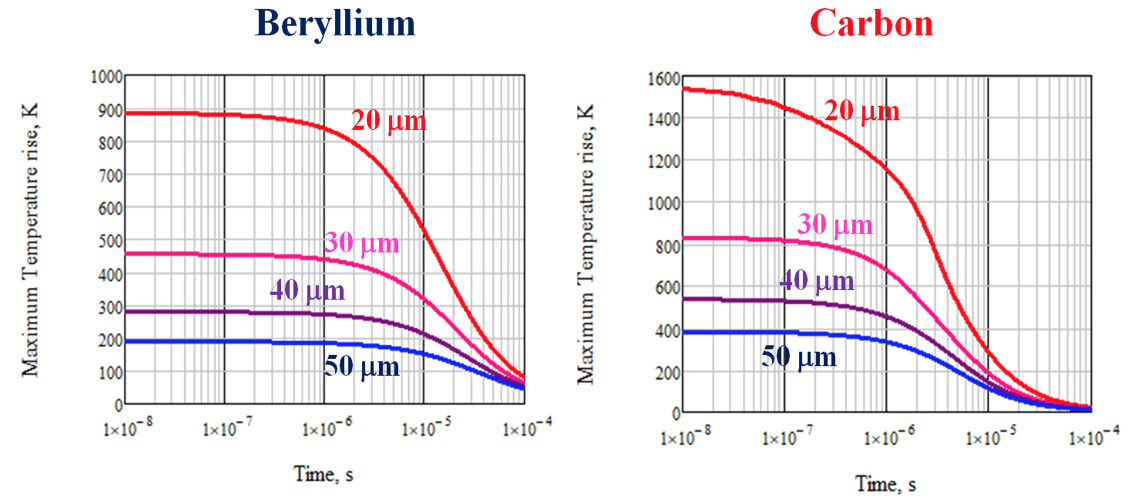}
    \caption{Temperature increase in the non-diffusion and cooling phases for Beryllium (3 mm) and Carbon (1 mm) by varying the dimensions of the incident beam spot.}
    \label{fig:temp}
\end{figure}

Both temperature rise and thermal shock are related to the beam size on target. 
For a given material the lower limit on the beam size is obtained when there 
is no pile-up of bunches on the same target position.
For this reason both the target and the positron beam have to be movable.
Fast moving targets can be obtained with rotating disks for solid targets or high velocity jets for liquids. 
A power deposition of about 30 kW is expected for a 0.3X$_0$ target. The target has to be therefore sliced in many thin targets
to easy the power removal.
Be and C composites/structures are in use and under study for low Z target and collimators in accelerators for high energy physics also because of the stringent vacuum requirements in such complexes that are not easy to fulfil with liquid targets.
Recently developed C based materials with excellent thermo-mechanical properties are under study for the LHC upgrade collimators~\cite{ipacT}. 
A 7.5 $\mu$s long beam pulse made of 288 bunches with 1.2$\times$10$^{11}$ protons per bunch, which is the full LHC injection batch extracted from SPS, has been used to test both C-based~\cite{ipacT} and Be-based~\cite{targetry} targets with maximum temperatures reaching 1000$^\circ$~C. 
Good results have been obtained with a beam spot of 0.3~mm$^{2}$ corresponding\footnote{for such thin light targets the energy deposition is largely dominated by ionization energy loss that is at first approximation similar for protons and $e^+$} to a LEMMA bunch intensity of  3$\times$10$^{11}$ particles,  on spot sizes as small as  $\sim$ 20 $\mu$m$^2$.
Good results have been obtained with a beam spot of $0.3\times0.3$~mm$^{2}$.
A first study of thermal behaviour has been performed both with Be and C. The target slice considered are 3 mm Be of 1 mm C.
For this purpose  Monte Carlo simulations have been performed with FLUKA both for Be (Fig. \ref{fig:pedd}a) and C  (Fig.\ref{fig:pedd}b). The figures show the heat deposited by a single bunch of $3\cdot10^{11}$ $e^+$ as a function of the radial distance from the center. The curves refer to different beam spot sizes a respectively for 10, 20, 30, 40, 50, 140, and 300 $\mu$m.
The spatial and temporal distribution of the thermal field has been calculated  using the Fourier heat transfer 
from the heat density deposited  taking into account  the dependence on temperature of the thermal parameters of the material.
A Finite-difference time-domain method (FDTD) code has been developed for the evaluation of the temperature gradient on the target and the timing of heat diffusion on the latter. 
This procedure has been first applied to the case of a single bunch and allowed to identify three temporal regions for the temperature field.
In the first temporal region, during the interaction between the $e^+$ and the target, there is a rapid temperature increase (t $<$ 10ps). After the pulse (second temporal region) the heat initially does not diffuse and remains confined to the area of interaction (order of $\mu$s); after a third phase begins where heat diffuses radially to the colder regions of the disk. 
Considering the variability with temperature of the thermal parameters and spot sizes ranging from 20 µm to 50 µm we obtain for a single bunch the temporal trends of temperature on target represented in Fig.\ref{fig:temp}.
The evaluation of the spatial and temporal temperature gradients produced on the target is very useful for estimating the thermo-mechanical stress induced by a single bunch or a sequence of bunches. In order to obtain the working temperature of the target in steady state and the trend of the residual temperature, we used a simple model based on the energy balance between the energy deposited by the sequences of the positron pulses and the energy dissipated by radiation only.
We consider the ILC Positron Source Target rotation system\footnote{the LEMMA target is much lighter than the ILC one and in principle higher velocities could be envisaged}. 
The equilibrium working temperature has been determined for a structured target in the shape of a circular crown and arranged on a rotating support at speed of  100 m/s with diameter of 1 m. The increase in the working temperature, with respect to the room temperature (298 K), of the target in steady state, assuming an external radius of 55 cm and an inner radius of 45 cm  is 78 K and 36 K respectively for a bunch train of 1000 bunches and a repetition frequency of 10 Hz.
This preliminary study shows that Be or C target with an ILC like rotating system could be used for the LEMMA target: both Be and C have been tested at the expected the expected peak temperatures and the expected steady state temperature is  below the one expected for the ILC target.
Liquid jet target are also a viable option for LEMMA. A very interesting option is represented by the Hydrogen pellet/spaghetti target. Only first vacuum consideration have been done in this case showing that with 1000 l/s turbo pumps there will be average H$_2$ pressure of 2.7 mbar. A very high vacuum impedance (small beam pipe tube) of the order of few mm of diameter and meters of length is needed to separate this bad vacuum region from the other parts of the accelerator. In addition the peak value during oh the H$_2$ pressure occurring during the positron train passage need to be carefully evaluated as well as the pressure time evolution.  

\section{TARGET LINES}
\begin{figure}[ht]
    \centering
	\includegraphics[width=\columnwidth]{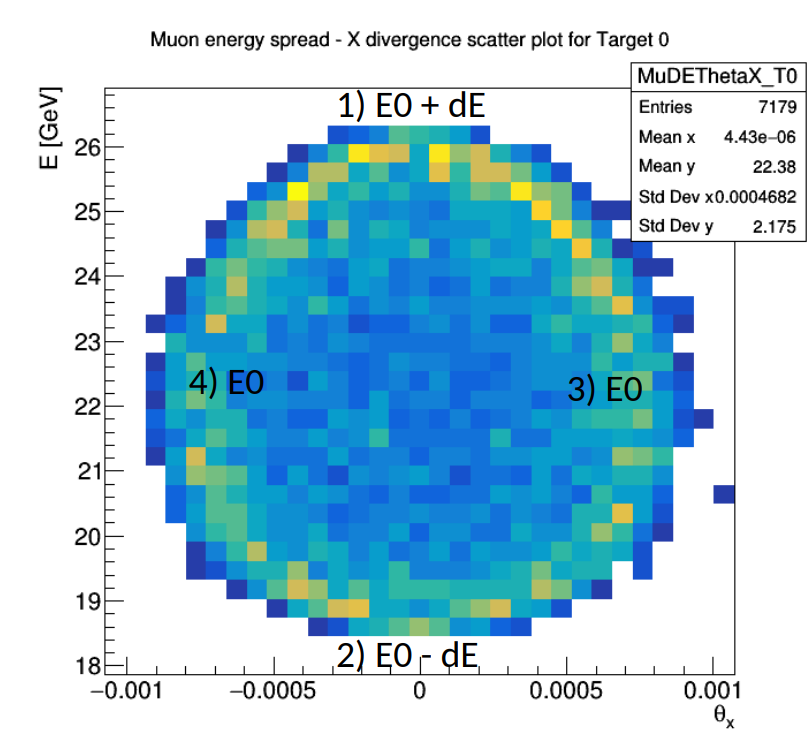}
    \caption{Energy vs Angle of muons produced from a positron beam with small energy spread and divergence.}
    \label{fig:muon_Evsangle}
\end{figure}

For the previous LEMMA scheme \cite{PRAB} simulations in Geant~\cite{geant} and AT~\cite{AT} have been performed to study the effect on the positron beam of a target inserted in the PR and the distribution of the produced muons in the phase space~\cite{Boscolo:IPAC2018-MOPMF087}. We recall in Fig.~\ref{fig:muon_Evsangle} the muon energy as a function of the production angle, for muons produced by a 45~GeV positron beam impinging on a Berillium target with small energy spread and divergence.\par
The muon energy and angle of production depend one on the other due to kinematics. Given that the positron beam divergence ($\sigma'_{e+}$) and energy spread are small, the maximum values for the angle and energy distribution of the out-going muons depend on the positron beam nominal energy.\par
We can calculate approximatively the r.m.s. emittance of the muon beam as $\epsilon_\mu = \sigma_{e+}\cdot\sigma'_{\mu}(E_{oe+})$, where $\sigma_{e+}$ is the positron beam size, and $\sigma'_{\mu}(E_{oe+})$ is the muon divergence depending on the positron beam energy $E_{oe+}$.  For a 45~GeV positron beam, we have $\epsilon_\mu \approx 0.5$~mrad $\cdot \sigma_{e+}$.\par
This scheme showed both a large average energy deposition and a large Peak Energy Density Distribution (PEDD) on the target. To improve these parameters a new option has been studied, where the positron beam is extracted from the PR and impinges on one or more targets outside the ring, in a so-called Target Line. The idea is that to increase the number of produced muons we could multiply the number of  Target Lines, provided a ``fresh'' positron beam is used in each one. In the following a description of the studies performed for this option is summarized.

\subsection{Multiple IPs, multiple targets}
\begin{figure*}[ht]
    \centering
	\includegraphics[width=2\columnwidth]{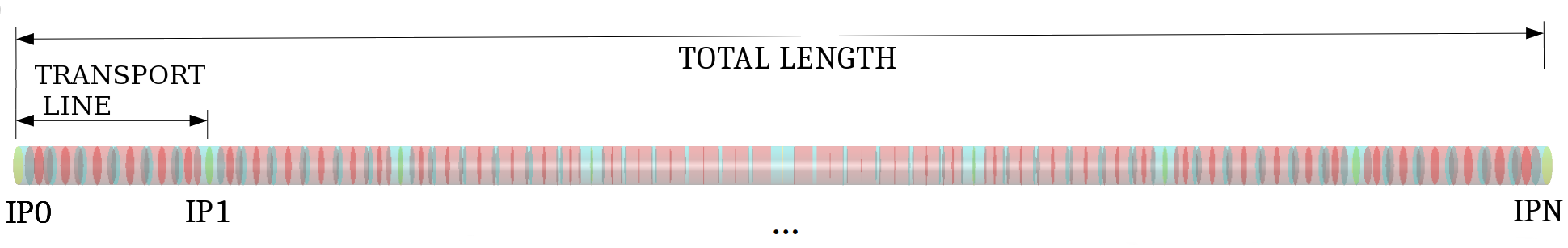}
    \caption{Muon and positron beam transport through a common line with targets in multiple IPs.}
    \label{fig:transport}
\end{figure*}

The first layout studied has been the collision of the positron beam with several targets on a line. In addition to splitting the target in several slices on one IP, the multiple IP concept consists in the separation of the targets by a transport line where magnets are common to the three beams ($e^+$, $\mu^+$ and $\mu^-$). This transport line should focus the beams at each IP to achieve the production of new muons with minimal growth to the final beam emittance. A sketch diagram is shown in Fig.~\ref{fig:transport}.\par
Several constraints in the design had to be balanced. First, the length should be as small as possible in order to minimize muon decay issues. Secondly, focalizing three beams at different energies imposes constraints on the minimum number of elements in the line. Then, chromaticity cannot be corrected with dipoles+sextupoles because this would split the three beams, therefore, other methods should be used to mitigate the chromatic effect. Moreover, we will need a minimum amount of space between IPs and closest quadrupoles to accommodate the targets. Lastly, the beam divergence has to be larger than the effect of multiple scattering to mitigate the emittance growth.\par
\begin{figure}[ht]
    \centering
	\includegraphics[width=\columnwidth]{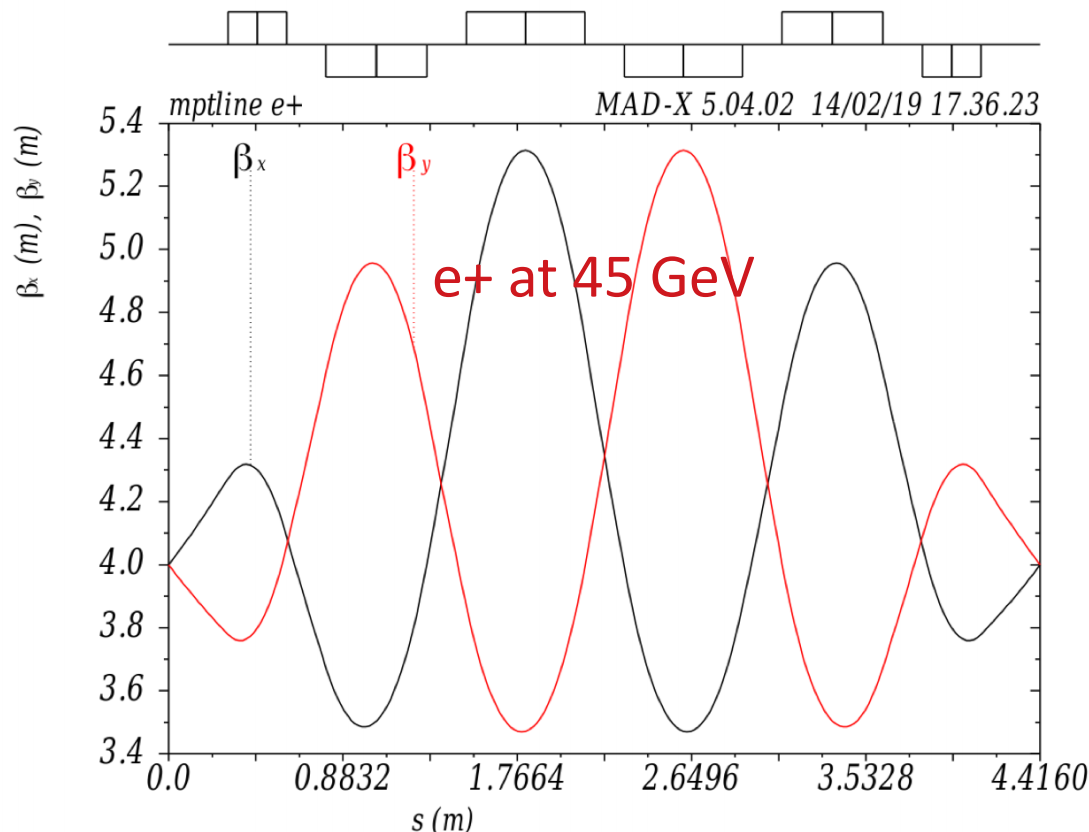}
    \caption{Transport line optics at 45~GeV.}
    \label{fig:optics45GeV}
\end{figure}
\begin{figure}[ht]
    \centering
	\includegraphics[width=\columnwidth]{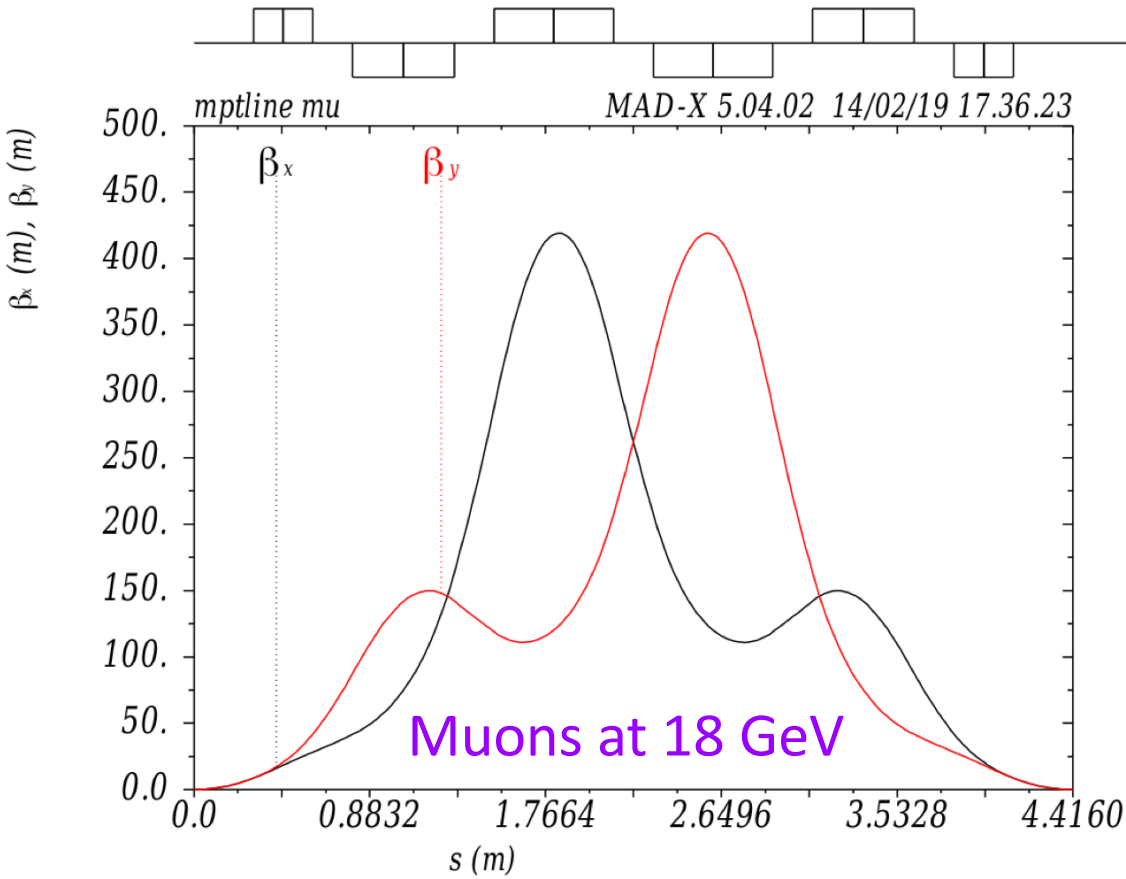}
    \caption{Transport line optics at 18~GeV.}
    \label{fig:optics18GeV}
\end{figure}

Fixing the distance from IP to quads to 30~cm, we present the best lattice design. It is less than 5~m long, with quadrupole magnet gradients at 200~T/m, 1~cm of aperture radius, separated by drift spaces of about 20~cm. Two triplets are used to focus the beams at 45~GeV and 18~GeV on both transverse planes. These triplets are put in asymmetry in order to partially cancel chromaticity at 45~GeV as in the apochromatic design~\cite{PhysRevAccelBeams.19.071002}. Optics functions calculated in MAD-X~\cite{MADX} at both energies are shown in Figs.~\ref{fig:optics45GeV} and~\ref{fig:optics18GeV}.
\\
Chromaticity at the muon beam energy is not cancelled, therefore the muon beam shows a rapid emittance growth as shown in Fig.~\ref{fig:c1}, magenta line. Between the first and second IP, the emittance grows because of the combination of chromaticity and large energy spread of the muon beam, as they are produced between 18.5~GeV and 26~GeV, i.e.~$\pm18\%$ energy spread.
The final achieved emittance is just below 200~nm, giving an important contribution larger than a factor two to the initial emittance. Several additional lattice optics configurations where tried to minimize the effect of chromaticity at the expense of lower energy acceptance. 

In summary, the studies on this configuration showed that, since the chromaticity for the muon beam is not cancelled in the TL, the emittance of the muon beam is largely increased. Therefore, a new layout with a single IP, with several targets very close to each other, has been studied.
\subsection{Single IP, ten targets}
\begin{figure}[ht]
    \centering
	\includegraphics[width=\columnwidth]{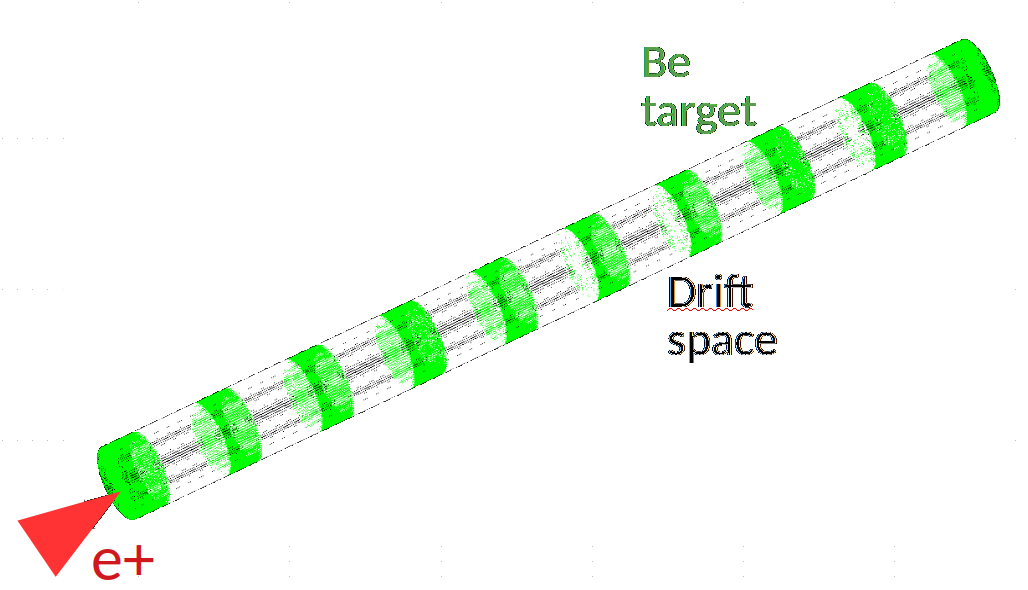}
    \caption{Single IP with multiple targets.}
    \label{fig:SingleIP}
\end{figure}

\begin{figure}[ht]
    \centering
	\includegraphics[width=\columnwidth]{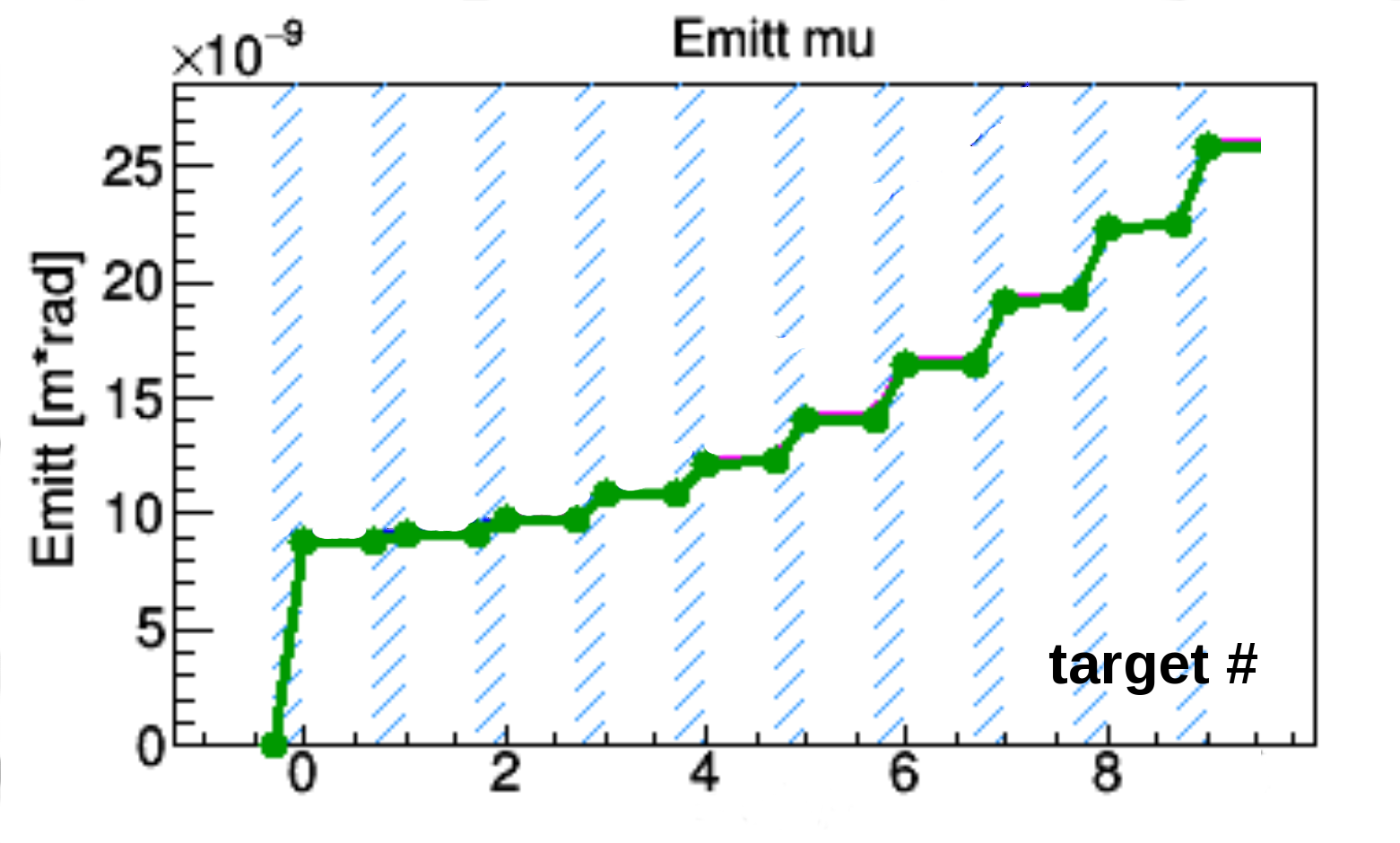}
    \caption{Muon beam emittance when crossing 0.3~R.L. of Beryllium divided in 10 pieces separated by 2~cm. The positron beam size is 20~$\mu$m and emittance of 0.07~nm.}
    \label{fig:MuonEmit10nm}
\end{figure}

\begin{figure}[ht]
    \centering
	\includegraphics[width=\columnwidth]{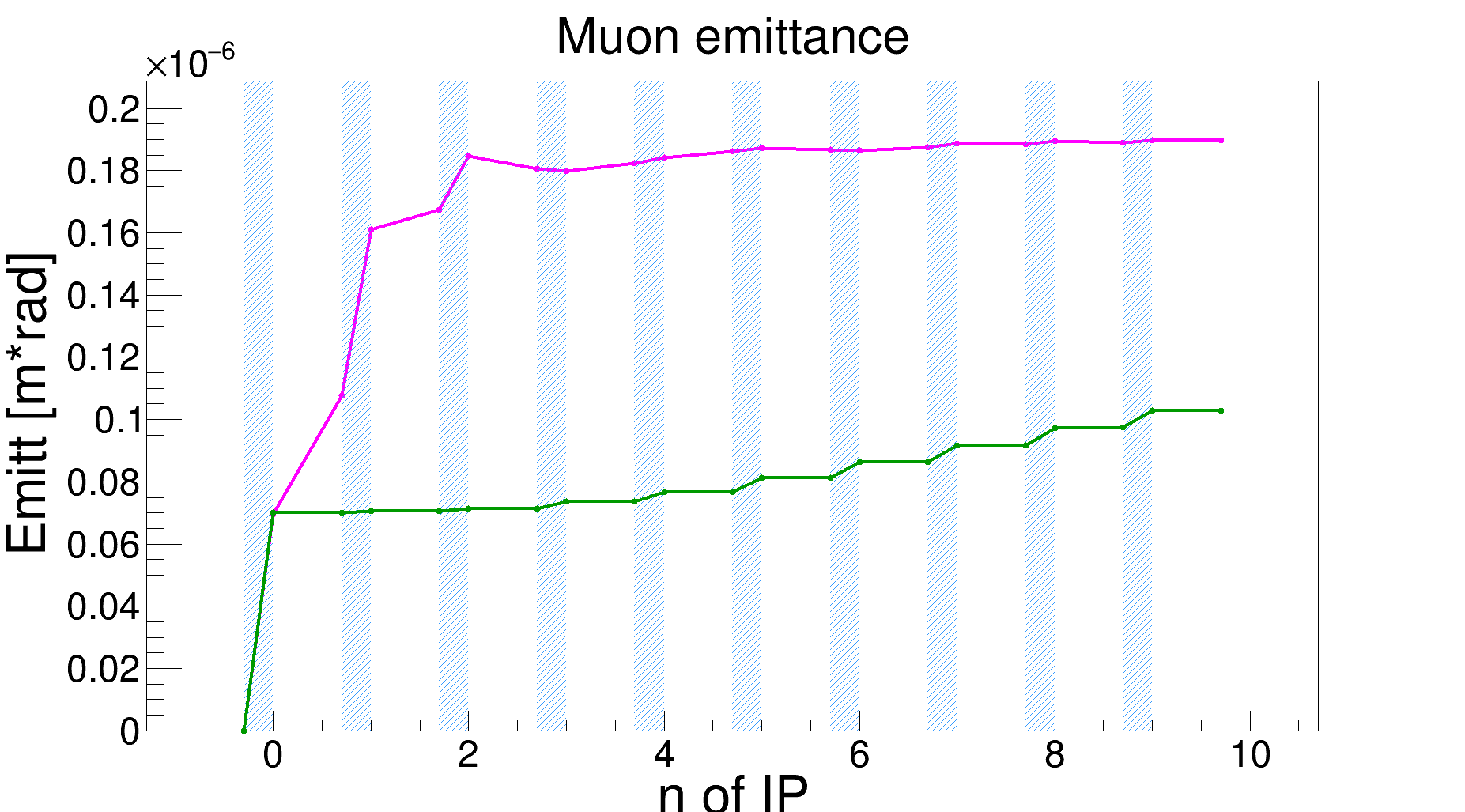}
    \caption{Comparison of $\mu$ emittance growth in the Multiple (magenta) and Single (green) IP schemes as a function of the target number (0 to 9). The $e^+$ beam size is 150~$\mu$m in both cases.}
    \label{fig:c1}
\end{figure}

In order to reduce the emittance of the muon beam, a design with just one IP has been exploited. This single IP consists in having one region where $e^+$ collide with several targets that have been distributed in slices aligned with the $e^+$ beam and separated by very small drifts in order to give space for power dissipation. Figure~\ref{fig:SingleIP} shows schematically the region under discussion.
We estimated two cases: the first one from a positron beam size at the first target of $\sigma_{e+}=20$~$\mu$m and emittance of $\epsilon_{e+}=70$~pm (Fig.~\ref{fig:MuonEmit10nm}). The 10~nm muon emittance in the first target is given by the beam size of impinging $e^+$ as explained before, while, the emittance grows to 26~nm in the nine consecutive targets because of the muon beam multiple scattering with the targets. A second, more conservative case, from a $e^+$ beam size $\sigma_{e+}=150$~$\mu$m and emittance of $\epsilon_{e+}=6$~nm gives a muon beam emittance of 70~nm (see Fig.~\ref{fig:c1}, green line) and grows up to 110~nm.

A comparison of muon emittance growth for the multiple versus the single IP design is shown in Fig.~\ref{fig:c1} for the same $e^+$ beam spot on the target (150~$\mu$m). Of course the smaller the $\sigma_{e+}$ at the targets, the smaller will be the ``absolute'' $\mu$ emittance growth. At present, different $e^+$ beam spots on the target are being studied, as well as different target materials, since this parameter is crucial both for the $\mu$ emittance and for the amount of PEDD and temperature rise of the target.
\section{MUON PRODUCTION EFFICIENCY}
\begin{figure}[ht]
    \centering
	\includegraphics[width=\columnwidth]{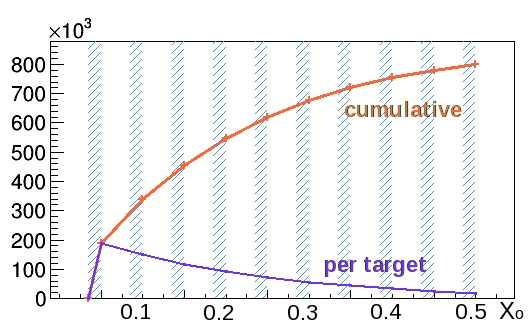}
    \caption{Number of muon pairs produced by 5$\times10^{11}$~e$^+$ vs ten Beryllium targets of 5\% radiation length each. The positron beam population above 43.7~GeV is reduced by bremsstrahlung.}
    \label{fig:muon_number}
\end{figure}

Using a fast Monte-Carlo validated with Geant4, named MUFASA~\cite{ciarma}, we have calculated the number of muon pairs produced by 5$\times10^{11}$~e$^+$ at 45~GeV, impinging on ten Beryllium targets of 5\% of a radiation length each, see Fig.~\ref{fig:muon_number}.\par
The positron beam loses energy due to the effect of bremsstrahlung in the target, and although the positron population is not reduced, the number of produced muons decreases because less particles remain above the muon production threshold at 43.7~GeV.
As almost 90\% of the muons are produced in the first 6 targets, equivalent to 0.3~X$_o$ or 10.6~cm, we will use it to calculate the target \emph{production efficiency}, i.e. the ratio of muon pairs produced by $e^+$ impinging on a target, $\mu/e^+$.\par
Therefore, the efficiency of a 0.3~X$_o$  Beryllium target is $\mathbf{1.3\times10^{-6}\mu/e^+}$ ($674\times10^{3}\mu/5\times10^{11}e^+$).\par
\begin{figure}[ht]
    \centering
	\includegraphics[width=\columnwidth]{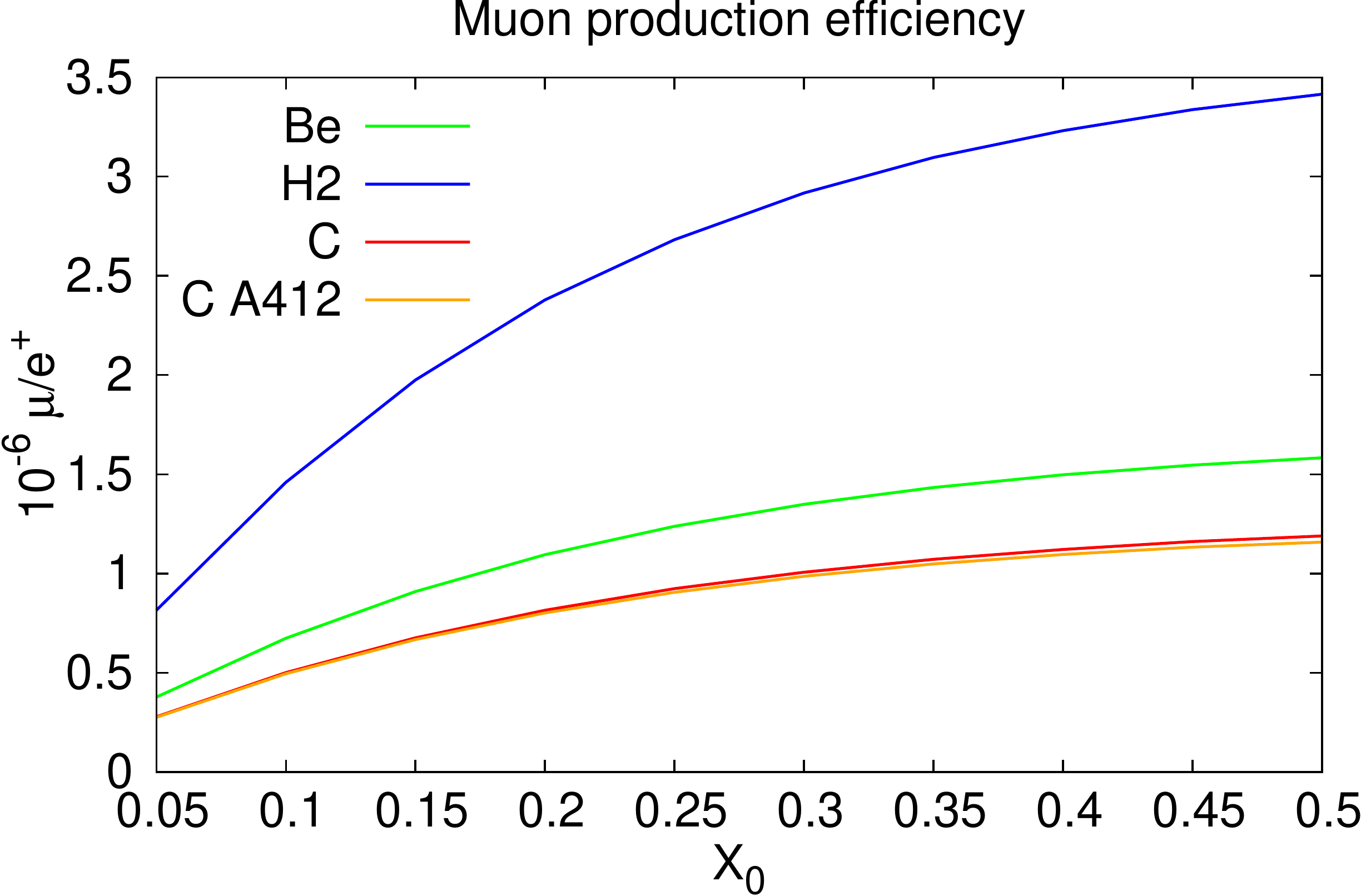}
    \caption{Muon pair production efficiency versus material thickness in radiation length units.}
    \label{fig:materials}
\end{figure}

The Beryllium efficiency was compared with two possible Carbon materials because of the higher resistance to thermal stress. Table~\ref{tab:Ceff} shows the results. We remark that Carbon composites would reduce the muon production efficiency by about 25\%. Liquid H$_2$, with a density of 0.07~g/cm$^3$, would double the efficiency in the same fraction of radiation length,  which could be an advantage because it minimizes emittance growth from multiple scattering; the possibility to use it as target for our needs is planned as future R\&D study. In addition, Fig.~\ref{fig:materials} shows the efficiency of muon pair production as a function of the target thickness for these materials.\par
\begin{table}[ht]
    \centering
    \caption{Muon production efficiency for Beryllium, two Carbon composites and liquid Hydrogen. Density and length in m and X$_0$ are included.}
    \label{tab:Ceff}
    \begin{tabular}{lcccc}
        Material & Density & \multicolumn{2}{c}{Length} &  eff  \\
        & [g/cm$^3$]&[m] &  [X$_0$]  & [$10^{-6}\mu/e^+$]\\\hline
        Be & 1.85 & 0.106 & 0.3 &  1.3\\
        C      & 2.27 & 0.057  & 0.3 &  1.0 \\
        C A412 & 1.7\;\;  & 0.075 & 0.3 & 1.0 \\
        H$_2$  & 0.07 & 2.664 & 0.3 &  2.9
    \end{tabular}
\end{table}

\subsection{From positron beam at 48~GeV}
\begin{figure}[ht]
    \centering
	\includegraphics[width=\columnwidth]{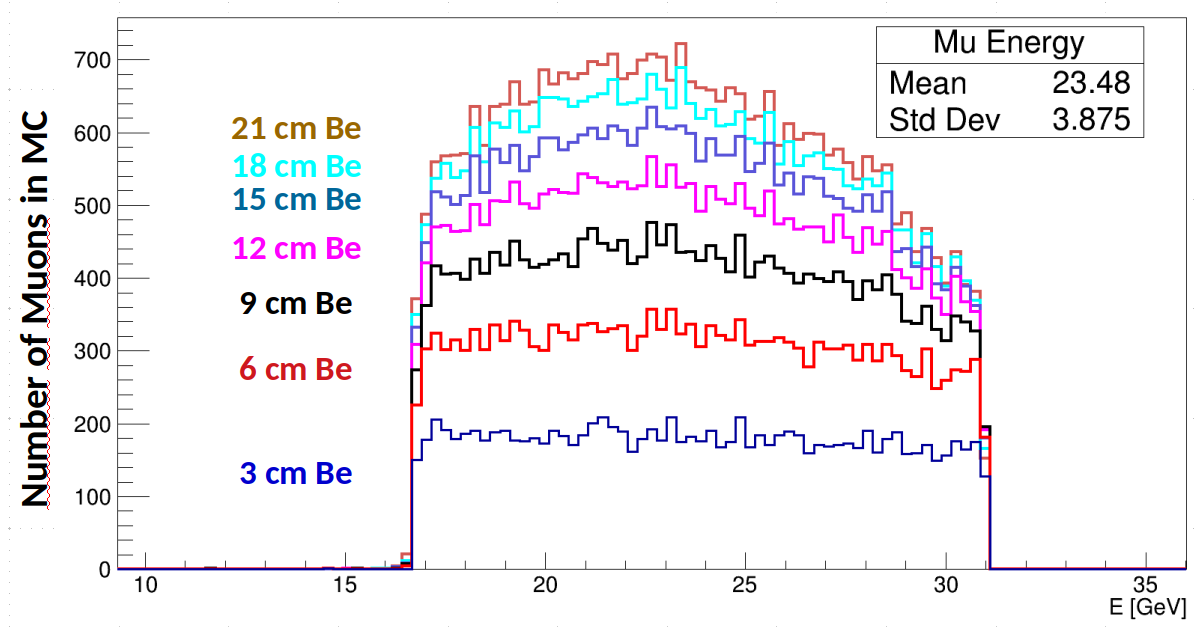}
    \caption{Muons produced in Monte Carlo simulation from a 48~GeV positron beam.}
    \label{fig:energyscan}
\end{figure}

Bremstrahlung in the target reduces the positron beam population above 43.7~GeV, the production energy threshold. In particular, starting with a 45~GeV positron beam, most of the particles would cross a 0.3~X$_0$ target (10.6~cm of Beryllium), however, almost 2/3 of the population do not produce more muons. In order to explore the possibility to use thicker targets, we simulated the muon production from a 48~GeV positron beam using Geant, Cern--Root~\cite{cernroot} and MDISim~\cite{mdisim}. Figure~\ref{fig:energyscan} shows the result.\par
As expected, the effective production of muons extents to 20~cm of Beryllium, approximately 0.57~X$_0$ but the energy spread of the muon beam increases from $\pm1.8$~GeV to $\pm3.9$~GeV. 
\begin{figure}[!h]
\centering
\includegraphics[width=\columnwidth]{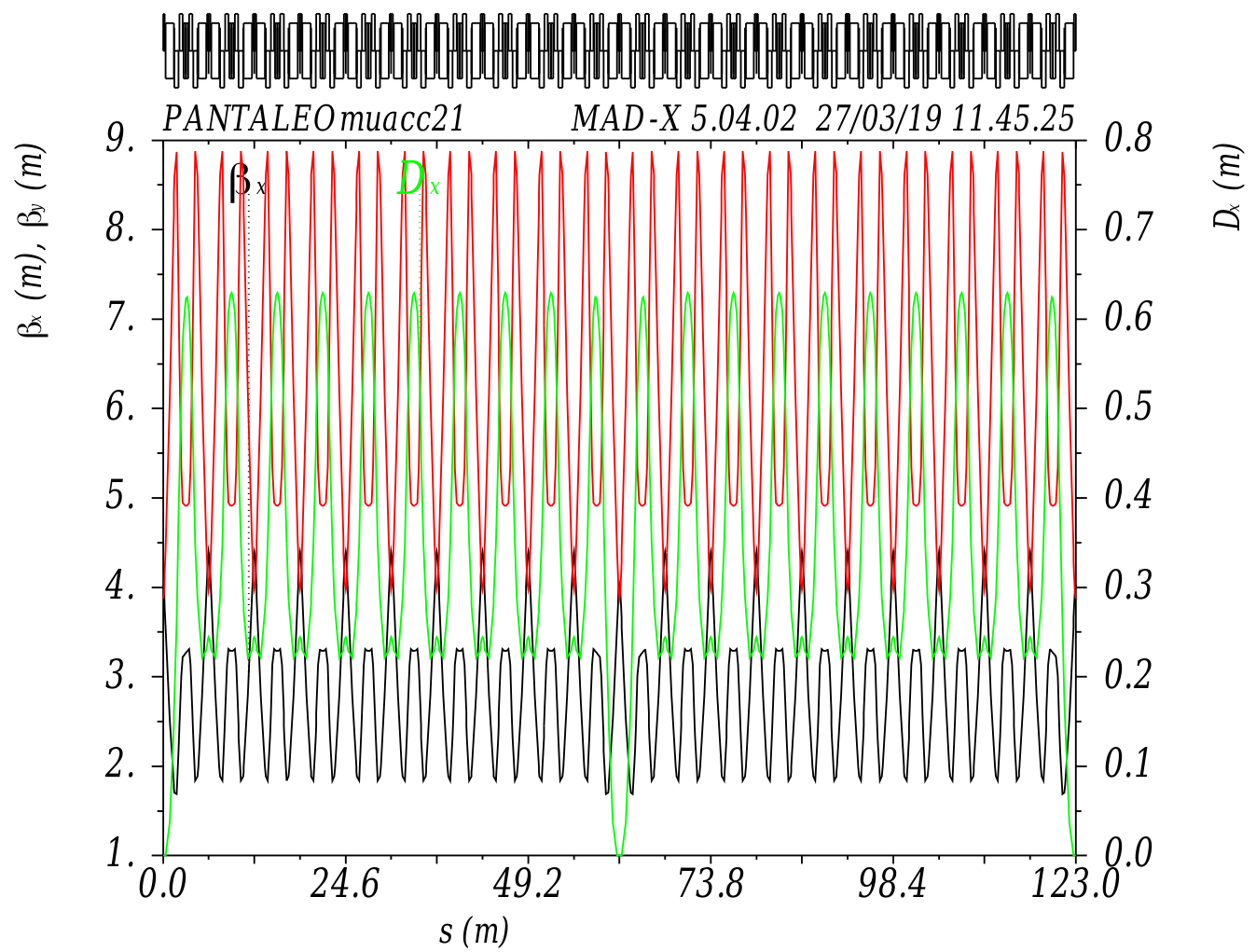}
\caption{Muon Accumulator: ~123~m lattice optical functions.}
\label{fig:acc}
\end{figure}
\section{Accumulator Rings}
The purpose of the Muon Accumulator Rings (MA) is to store the muons produced over several passages of the positron beam, therefore, their length must match the timing between positron bunch passages, i.e. new muons are created at the moment of passage of the stored muons therefore increasing the muon bunch intensity. On top of that, the MA must be short in order to complete a large number of turns before muons decay.\par
In the current scheme, muons are not injected but generated directly inside the ring. Therefore, the interaction region is common to a transport line for $e^+$ and two MA rings, one per muon species. A preliminary separation region is described in the following section, that could be also considered for the beam re-combination where particles direction are reversed.\par
Both designs are still not yet compatible, but, they point to the requirement of a zero-dispersion section in the MA rings to allocate space for the targets, and a minimal distance to separate the three beams.\par
\begin{figure}[!h]
\centering
\includegraphics[width=\columnwidth]{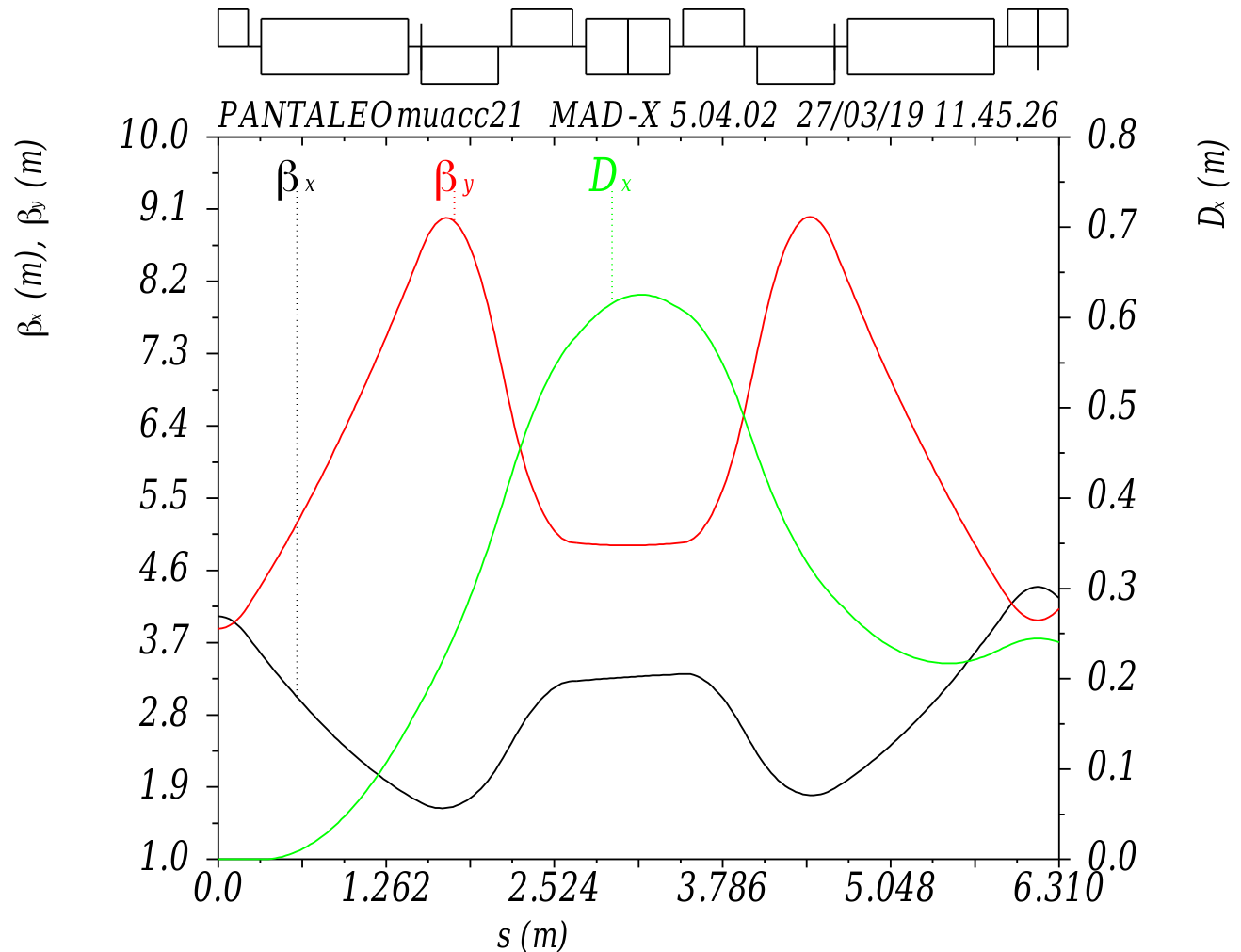}
\caption{Muon Accumulator: optical functions in zero dispersion region to allocate targets.}
\label{fig:accb}
\end{figure}

A preliminary optics design of the MA is shown in Figs.~\ref{fig:acc} and ~\ref{fig:accb}. The total length is 123~m, allowing the recirculation of the muon beam every 410~ns, and allowing to complete 1000 turns in one muon life time at 22.5~GeV.\par
It is a very compact design, composed by two zero-dispersion regions: one to allocate the target, and another to extract the muon beam at the end of the accumulation. In every cell, two long bending magnets of 16~T provide the closed curvature, while a short magnet at a similar field strength and opposite polarity is used to cancel part of the momentum compaction factor. The zero-dispersion region is a variation of this configuration where the magnets are at 18~T and -25~T respectively.\par
The space between magnets is 10~cm long, quadrupoles gradients are below 100~T/m and sextupole gradients are below 300~T/m. In addition, higher order components have been added to optimize dynamic aperture and momentum compaction factor.\par
At the moment of writing, the design does not reach the desired $\pm$20\% of energy acceptance required because of the muon beam production kinematics from a 45~GeV positron beam. Initial tracking studies allowed to estimate the momentum acceptance in less than $4\%$, but, with a very small momentum compaction factor for energies between $\pm7\%$ of the nominal 22.5~GeV.

\section{Beams Separation Region}
The three beams ($e^+$, $\mu^+$ and $\mu^-$), coming from the interaction of the positron beam and the targets, should be separated in a dedicated region to continue in the accelerator chain. The goal is to achieve the beam split with minimum particle losses.\par
Because of the large energy spread and divergence of the muon beams, this requires a careful study. In addition, the positron beam has been degraded due to breemstrahlung in the target, therefore, although most of the $e^+$ enter the separation region, only a fraction will pass through to the next stages.\par
At the exit of the separation region we expect to have three independent beam lines, one per particle type, with enough separation to allocate independent optics elements to transport the positron beam into the recycling line, and the muon beams into the MA rings.\par
\begin{figure}[!h]
\centering
\includegraphics[width=0.8\columnwidth]{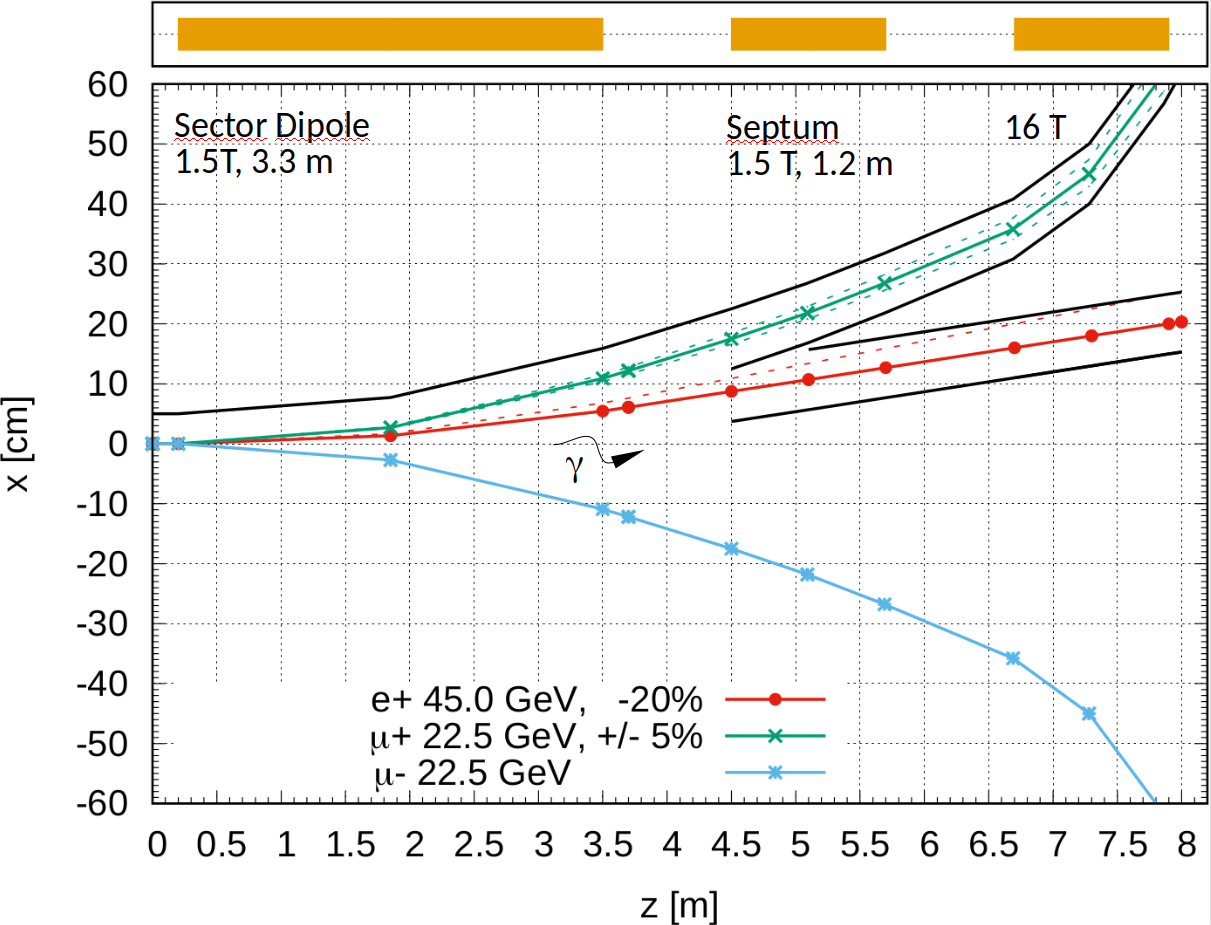}
\caption{Beams separation region.}
\label{fig:beamseparation}
\end{figure}

A very preliminary conceptual diagram is shown in Fig.~\ref{fig:beamseparation}. It is composed by two dipoles separated by 1~m. First, a sector dipole of 1.5~T, 3.3~m long starts the beam separation before they reach the septum magnet of 1.5~T, 1.2~m long. In the diagram we also include a third dipole with a magnetic field of 16~T that would correspond to the first dipole in the MA rings.\par
The magnetic field of the first magnet has been kept below 2~T to avoid the use of superconductive magnets in a region close to targets radiation and secondary particles, and also because lower fields limit the photon radiation produced by the passage of the high energy positron beam. Synchrotron radiation photons are expected to be emitted by the positron beam with a critical energy of 2~MeV (in the hard X-ray range), and  in a very narrow cone of angle $\propto 1/\gamma_{rel}$, where $\gamma_{rel}$ is the relativistic energy constant of the positron beam. Therefore, some care must be taken downstream the sector dipole.\par
The energy lost by the positron bunch in this first magnets has been estimated to be 20~MeV, comparable to that coming from bremstrahlung in a fraction of a target radiation length.\par
The second magnet includes a septum of few milimiters in thickness, separating the beam lines in three parts : two regions with magnetic field for the muon beams, and one region without magnetic field for the positron beam. The diagram in Fig.~\ref{fig:beamseparation} shows only the separation of the two positive charges. A 5~cm aperture radius has been drawn around the positive muon and positron trajectories at nominal energy. This is to indicate and estimate the effect of a beam pipe in the particle losses.\par
At the entry point of the 16~T super conductive magnet, there will be a clearance of about 10~cm around the beam pipes in order to give space to the super conductive magnet. No further mechanical consideration has been set for the moment, and this will need to be checked in further studies.\par
The current configuration allows the passage of muons with $\pm$5\% of energy spread, meaning that it will need to be redesigned to accept $\pm20\%$ of energy spread. \par
Positrons with energy above 36~GeV, i.e. 20\% energy loss, would be able to cross the 8~m long separation region. Table~\ref{tab:percentagealive} shows the percentage of $e^+$ entering and exiting the separation region for several target thicknesses distributed in one or several interaction points.\par
\begin{table}[h]
    \centering
    \begin{tabular}{ccc}
         Target Thickness & Entry & Exit \\
         \% R.L. (Distribution) & \% & \% \\\hline
         $\;\;$0.9 (~1~IP$\times$3~mm)  & 100 & 99 \\
         $\;\;$8.5 (10~IP$\times$3~mm) & 98 & 87 \\
         14.2 (10~IP$\times$5~mm) & 95 & 80 \\
         28.4 (10~IP$\times$10~mm) & 88 & 61
    \end{tabular}
    \caption{Percentage of $e^+$ entering and exiting the separation region for different Beryllium target thickness distributed in one of several IPs.}
    \label{tab:percentagealive}
\end{table}
The particles losses reported here, in conjunction with the maximum possible positron population achievable per bunch, sets the rate of positron replacement, either by injection or regeneration, that the entire positron cycle will require. Therefore, the lower the losses the lower the positron source requirements. On the opposite side, there is an advantage in putting a larger fraction of target material to generate more muon pairs per passage.  A balance between the two cases should be found, in addition to the re-conception of the separation region to reduce the shown losses.

\section{FUTURE R\&D}
To increase the muon beams quality, and consequently the final luminosity, in the proposed scheme different proposals are conceivable if a solid R\&D program could demonstrate their technical feasibility. Improvements in the technical solution could enhance not only each system performances, but in some case the global efficiency of the full muon source complex. We will briefly summarize hereafter the main possible directions for the R\&D programs, their correlation with the source parameters and their functional relationship with the final luminosity.
\subsection{MUON TARGETS}
As already mentioned one of the most important parameter to increase the muon bunch population is the possibility to produce the maximum number of $\mu^+$ $\mu^-$ pairs in a single positron bunch passage, up to the limit of its energy and energy spread deterioration that fix the limit to the use of a ``fresh'' bunch. To maximize this parameter, the Be and C targets were considered since, thanks to their low Z, they present a lower Z(Z+1) dependent bremsstrahlung effect. At present the schemes takes into account that an integrated 0.3X$_0$ target thickness is suitable for a single positron bunch passage. A very important development should represent the possible use of Hydrogen targets that, mixed with the multi-IP lines, will improve the integrated thickness reducing the number of passages and so increasing the ration of ``fresh'' bunches/passages. This will have a linear dependence on the muon per bunch number, and so a quadratic increase of the final luminosity. Taking into account a simple scaling with Z we expect a factor 15 in increase of the luminosity.
\subsection{POSITRON SOURCES}
One of the main limit in the source repetition frequency is the physical constraint imposed by the $e^+$ source given by the required $e^+$ flux, the required cooling and the thermos-mechanical stress on the target. In this framework a very interesting development is represented by the use of rotating target as already conceived for the ILC. Different schemes at a $f_{rep}$ of 50-100 Hz should be implemented in case that high technology targets and high efficiency $e^+$ source should deliver $e^+$ rate higher than $10^{16}$ $e^+$/sec. This has to take into account also the possibility to develop immersed $e^+$ capture systems with very high peak B Field in the AMD (20 T in the MAPS scheme) and in the capture solenoid. A very large energy spread of the damping ring will also increase the efficiency of the $e^+$ source also if this has to be carefully harmonized with the cooling time. An increase in the efficiency of the $e^+$ source, and so of the repetition rate of a factor 5-10, will have a linear dependence on the luminosity.
\subsection{POSITRON RING}  	
In all the illustrated schemes one of the imposed limit is given by the achievable current in the PR. This is mainly due to the beam instabilities and to the synchrotron power budget. In scheme I it was mentioned the possibility to work with the PR at a reduced energy, so drastically reducing the emitted power, and to accelerate and decelerate the $e^+$ beam respectively before and after the $e^+$ production in a push pull configuration. To implement this scheme, it is necessary to develop high gradient accelerating systems that can work at a very high value of the pulse current, typically 250 mA, in the 410 $\mu$sec allowed for the muon production cycle. This should introduce the possibility to increase the ring current of a factor 3-4 so increasing the number of available ``fresh'' bunches. In this case we should have more passages to produce the muons so increasing the number of muons per bunch so having a quadratic effect on the luminosity.
\subsection{MUON COOLING} 
The LEMMA scheme, despite of the low production cross section, introduces two main advantages in the source: a reduced emittance at the production and a higher production energy resulting in a longer muon lifetime. So, also if the former suggest the possibility to avoid a cooling phase, the second allow for enough time to introduce also a moderate cooling mechanism to further reduce the production emittance. Different evaluations were done in the past for the cooling efficiency given by stochastic cooling \cite{Ruggiero}, optical stochastic cooling \cite{Opt1,Opt2}, crystal cooling \cite{crycool}. A full revaluation of these mechanisms associated to high energy, low emittance and bunch current, long lifetime muon bunches should be produced and R\&D programs proposed, targeting an emittance reduction of 1-2 order of magnitude that will linearly impact on the final luminosity.

\section{CONCLUSIONS}
A conceptual accelerator design phase is necessary to evaluate the LEMMA scheme feasibility and to address the possible R\&D program to be pursued.
The fundamental part of this design is represented by the muon source complex. In this framework, three different schemes, taking into account the full muon production cycle, have been described. 
The different sub--systems operating in the schemes were analyzed, some simulations evaluated their final performances. 
This activity allowed to assess the conceptual feasibility of the LEMMA scheme and to identify the R\&D path and the design development directions to be followed to achieve the required collider luminosity.

\section{ACKNOWLEDGEMENTS}
G.Cesarini acknowledges the support of the ERC Ideas Consolidator Grant No. 615089 CRYSBEAM.

\flushcolsend
%
%
\ifboolexpr{bool{jacowbiblatex}}%
	{\printbibliography}%
	{%
	

} 
%
%


\end{document}